\documentclass[aps,pra,reprint,amssymb,superscriptaddress]{revtex4-1}

\usepackage{amsmath,amssymb,color}
\usepackage{bm}
\usepackage{hyperref}
\usepackage{graphicx}

\begin{document}

\title{Angular Momentum Conservation and Phonon Spin in Magnetic Insulators}

\author{Andreas R\"{u}ckriegel}
\affiliation{Institute for Theoretical Physics and Center for Extreme Matter and Emergent Phenomena,
Utrecht University, Leuvenlaan 4, 3584 CE Utrecht, The Netherlands}

\author{Simon Streib}
\affiliation{Kavli Institute of NanoScience, Delft University of Technology, Lorentzweg 1, 2628 CJ Delft, The Netherlands}

\author{Gerrit E.~W.~Bauer}
\affiliation{Kavli Institute of NanoScience, Delft University of Technology, Lorentzweg 1, 2628 CJ Delft, The Netherlands}
\affiliation{Institute for Materials Research \& WPI-AIMR \& CSRN, Tohoku University, Sendai 980-8577, Japan}

\author{Rembert A. Duine}
\affiliation{Institute for Theoretical Physics and Center for Extreme Matter and Emergent Phenomena,
Utrecht University, Leuvenlaan 4, 3584 CE Utrecht, The Netherlands}
\affiliation{Center for Quantum Spintronics, Department of Physics, Norwegian University of Science and Technology, 
NO-7491 Trondheim, Norway}
\affiliation{Department of Applied Physics, Eindhoven University of Technology,
P.O. Box 513, 5600 MB Eindhoven, The Netherlands}

%\date{\today}
\date{March 2, 2020}

\begin{abstract}
We develop a microscopic theory of spin-lattice interactions in magnetic
insulators,  separating rigid-body rotations and the internal angular
momentum, or spin, of the phonons, while conserving the total angular
momentum. In the low-energy limit, the microscopic couplings are mapped onto
experimentally accessible magnetoelastic constants. We show that the transient
phonon spin contribution of the excited system can dominate over the magnon spin,
leading to nontrivial Einstein-de Haas physics.
\end{abstract}

\maketitle

\section{Introduction}

The discovery of the spin Seebeck effect led to renewed interest in
spin-lattice interactions in magnetic insulators \cite{Uchida2010,Xiao2010},
i.e., the spin current generation by a temperature gradient, which is strongly
affected by lattice vibrations
\cite{Jaworski2011,Uchida2011,Kikkawa2016,Schmidt2018,Streib2019}. The
spin-lattice interaction is also responsible for the dynamics of the angular
momentum transfer between the magnetic order and the underlying crystal
lattice that supports both rigid-body dynamics and lattice vibrations, i.e.,
phonons. In the Einstein-de Haas \cite{Einstein1915} and Barnett effects
\cite{Barnett1915}, a change of magnetization induces a global rotation and
vice versa. While both effects have been discovered more than a century ago,
their dynamics and the underlying microscopic mechanisms are still under
debate \cite{Jafaar2009,Garanin2015,Dornes2019,Mentink2019}.

In 1962 Vonsovskii and Svirskii \cite{Vonsovskii1962,Levine1962} suggested that circularly polarized
transverse phonons can carry angular momentum, analogous to the spin of
circularly polarized photons. This prediction was confirmed recently by
Holanda \textit{et al.} \cite{Holanda2018} by Brillouin light scattering on
magnetic films in magnetic field gradients that exposed spin-coherent
magnon-phonon conversion. Ultrafast demagnetization experiments can be
explained only by the transfer of spin from the magnetic system to the lattice
on subpicosecond time scales in the form of transverse phonons
\cite{Dornes2019}. Theories address the phonon spin induced by Raman
spin-phonon interactions \cite{Zhang2014}, by the relaxation of magnetic
impurities \cite{Garanin2015,Nakane2018}, by temperature gradients in magnets
with broken inversion symmetry \cite{Hamada2018}, and phonon spin pumping into
nonmagnetic contacts by ferromagnetic resonance dynamics \cite{Streib2018}.
The phonon Zeeman effect has also been considered \cite{Juraschek2019}. The
quantum dynamics of magnetic rigid rotors has been investigated recently in
the context of levitated nanoparticles
\cite{Rusconi2016,Prat-Camps2017,Rusconi2017,Ballestero2019}. Very recently
ferromagnetic resonance experiments have shown coherent magnon-phonon coupling
over millimeter distances \cite{An2019}.

Most theories of spin-lattice interactions do not conserve angular momentum
\cite{Eastman1966,Gurevich1996,Garanin2015}, thereby assuming the existence of
a sink for angular momentum. The magnetocrystalline anisotropy breaks the spin
rotational invariance by imposing a preferred magnetization direction relative
to the crystal lattice, while the lattice dynamics itself is described in
terms of spinless phonons. The resulting loss of angular momentum conservation
is justified when the spin-phonon Hamiltonian does not possess rotational
invariance in the first place \cite{Garanin2015}, e.g., by excluding rigid-body
dynamics of the lattice and/or by boundary conditions that break rotational
invariance. In the absence of such boundary conditions, the angular momentum
must be conserved in all spin-lattice interactions. Phenomenological theories
that address this issue
\cite{Tiersten1964,Tiersten1965,Melcher1972,Bonsall1976,Chudnovsky2005,Jafaar2009,Garanin2015,Nakane2018} 
incorporate infinitesimal lattice rotations due to phonons but do not
allow for global rigid-body dynamics and therefore cannot describe the physics
of the Einstein-de Haas and Barnett effects. Conversely, theories addressing
specifically Einstein-de Haas and Barnett effects usually disregard effects of
phonons beyond magnetization damping
\cite{Kovalev2005,Kovalev2007,Bretzel2009,Chudnovsky2015}.

Here we develop a theory of the coupled spin-lattice dynamics for sufficiently
large but finite particles of a magnetic insulator allowing for global
rotations. We proceed from a microscopic Hamiltonian that conserves the total
angular momentum. We carefully separate rigid-body dynamics and phonons, which
allows us to define a phonon spin and to obtain the mechanical torques exerted
by the magnetic order on the rigid-body and the phonon degrees of freedom (and
vice versa). The theory of magnetoelasticity is recovered as the low-energy
limit of our microscopic model in the body-fixed frame and thereby reconciled
with angular momentum conservation. We compute the nonequilibrium spin
dynamics of a bulk ferromagnet subject to heating and spin pumping in linear
response and find that in nonequilibrium the phonons carry finite spin, 
viz.~a momentum imbalance between the two circularly polarized transverse phonon
modes. We also show that the phonon spin can have nontrivial effects on the
rigid-body rotation; in particular, it can lead to an experimentally
observable, transient change in the sense of rotation during equilibration.

The rest of the paper is organized as follows: The spin-lattice Hamiltonian
and the decoupling of rigid-body dynamics and phonons is presented in
Sec.~\ref{sec:H}. The spin transfer in a bulk ferromagnet is studied in
Sec.~\ref{sec:T} within linear response theory. Section \ref{sec:conclusions}
contains a discussion of our results and concluding remarks. Expressions for
the total angular momentum operator in terms of the Euler angles of the
rigid-body rotation are presented in Appendix \ref{app:J}, while
Appendix \ref{sec:Commutation} details the phonon commutation relations in finite
systems. Finally, Appendix \ref{sec:Rates} addresses the relaxation rates computed
by linear response.

\section{Microscopic spin-lattice Hamiltonian}
\label{sec:H}
% Hamiltonian, total angular momentum conservation, mapping to magnetoelasticity
%
We address here a finite magnetic insulator of $N$ atoms (or clusters of
atoms) with masses $m_{i}$ and spin operators $\bm{S}_{i}$ governed by the
Hamiltonian
\begin{equation} \label{eq:H}
H=
\sum_{i=1}^{N} \frac{\bm{p}_i^2}{2 m_i} + V(\{\bm{r}_i\}) + V_{\rm ext}(\{\bm{r}_i\}) + H_S ,
\end{equation}
where $\bm{r}_{i}$ and $\bm{p}_{i}$ are the canonical position and momentum
operators of the $i$th atom, and the potential $V(\{\bm{r}_{i}\})$ is assumed
to be (Euclidean) invariant to translations and rotations of the whole body.
$V_{\mathrm{ext}}(\{\bm{r}_{i}\})$ accounts for external mechanical forces
acting on the body. Because of translational and rotational invariance, the
spin Hamiltonian $H_{S}$ depends only on $\bm{r}_{ij}=\bm{r}_{i}-\bm{r}_{j}$
and must be a scalar under simultaneous rotations of lattice and spin degrees
of freedom. Since $\bm{S}_{i}$ are pseudovectors and $\bm{r}_{ij}$ are
vectors, the spin-lattice interaction depends only on $r_{ij}=\sqrt
{\bm{r}_{ij}^{2}}$, $\bm{S}_{i}\cdot\bm{S}_{j}$ and even powers of
$\bm{S}_{i}\cdot\bm{r}_{ij}$ \footnote{Broken inversion symmetry would allow
Dzyaloshinskii-Moriya interactions of the form $\left(  \bm{r}_{il}%
\times\bm{r}_{jl}\right)  \cdot\left(  \bm{S}_{i}\times\bm{S}_{j}\right)  $.}.
Considering only pair interactions between two spins, we arrive at
$H_{S}=H_{\mathrm{Ex}}+H_{\mathrm{A}}+H_{\mathrm{Z}}$, with exchange
($\text{Ex}$), anisotropy ($\text{A}$), and Zeeman ($\text{Z}$) contributions
\begin{subequations} \label{eq:H_S}
\begin{align}
H_{\textrm{Ex}} =
& - \frac{1}{2} \sum_{\substack{ i,j=1 \\ i\neq j }}^N 
J(r_{ij}) \bm{S}_i \cdot \bm{S}_j ,
\label{eq:H_Ex}
\\
H_{\textrm{A}} =
& - \frac{1}{2} \sum_{i,j=1}^N 
K(r_{ij}) \left( \bm{S}_i \cdot \bm{r}_{ij} \right) \left( \bm{S}_j \cdot \bm{r}_{ij} \right) ,
\label{eq:H_A}
\\
H_{\textrm{Z}} =
& 
- \gamma \bm{B} \cdot \sum_{i=1}^N \bm{S}_i . 
\end{align}
\end{subequations}
Here $J(r_{ij})$ is an isotropic and $K(r_{ij})$ an anisotropic exchange
interaction, $\bm{B}$ is the external magnetic field, and $\gamma=g\mu
_{B}/\hbar$ is the (modulus of the) gyromagnetic ratio, defined in terms of
the $g$ factor and Bohr magneton $\mu_{B}$, and Planck's constant $\hbar$.
$H_{\mathrm{A}}$ encodes the interaction of the spins with the crystal lattice
or crystalline anisotropy, which in the long-wavelength limit reduces to the
conventional crystal field Hamiltonian in terms of anisotropy and
magnetoelastic constants \cite{Eastman1966,Gurevich1996} (see
Sec.~\ref{sec:Phenomenology}). The interactions $J(r_{ij})$ and $K(r_{ij})$ in
principle include dipolar interactions. A Hamiltonian of the form of
Eq.~(\ref{eq:H_S}) has been used recently to compute the relaxation of a
classical spin system \cite{Assmann2019}.

Ultimately, the origin of the Hamiltonian (\ref{eq:H_S}) lies in the spin-orbit coupling of the electrons: 
The anisotropic contribution (\ref{eq:H_A}) arises from the dynamical crystal field that affects the electronic orbitals and thereby the spin states,
whereas the position dependence of the exchange contribution (\ref{eq:H_Ex}) is due to the dependence of the electronic hopping integrals on the interatomic distances.
For ultrafast processes that occur on the timescales of the orbital motion,
a description of these intermediate, electronic stages of the spin-lattice coupling might be necessary;
however, this is beyond the scope of this work.

\subsection{Rigid-body rotations and phonon spin}
% outline ro-vib decoupling and indicate correction terms by order, equations of motion?

The Hamiltonian (\ref{eq:H}) commutes with and thereby conserves the total
angular momentum, i.e., the sum of intrinsic electron spin and mechanical
angular momentum. In a solid, the mechanical angular momentum arises from the
rotation of the rigid lattice and the internal phonon dynamics. We may
decouple the $6$ rigid-body and the $3N-6$ phonon degrees of freedom by the
following transformation:
\begin{equation} \label{eq:r_expansion}
\bm{r}_i = \bm{R}_{CM} + {\cal R}(\phi,\theta,\chi) 
\left[ \bm{R}_i + \sum_{n=1}^{3N-6} \frac{ \bm{f}_n(\bm{R}_i) }{\sqrt{m_i}} q_n \right] ,
\end{equation}
where $\bm{R}_{CM}$ is the center-of-mass position,
${\cal R}(\phi,\theta,\chi) = {\cal R}_z(\phi) {\cal R}_y(\theta) {\cal R}_x(\chi)$
is a three-dimensional rotation parametrized by the Euler angles $\phi$, $\theta$, and $\chi$
($\mathcal{R}_{\mu}(\alpha)$ denoting a rotation by an angle $\alpha$ around an axis $\bm{\hat{e}}_{\mu}$), 
$\bm{R}_{i}$ is the body-fixed equilibrium position of the $i$th particle, 
and the $q_{n}$ are the normal coordinates of the lattice, i.e., the phonons, 
with eigenfunctions $\bm{f}_{n}(\bm{R}_{i})$ that diagonalize the energy to second order in $q_{n}$:
\begin{equation}
V(\{\bm{r}_i\}) = V(\{\bm{R}_i\}) + \frac{1}{2} \sum_{n=1}^{3N-6} \omega_n^2 q_n^2 + {\cal O}(q^3) .
\end{equation}
In molecular physics this decoupling of rotations and vibrations is referred
to as Eckart convention \cite{Eckart1935,Louck1976,Littlejohn1997}. Neglecting
surface effects of the external forces on the phonons, we also have
$V_{\mathrm{ext}}(\{\bm{r}_{i}\}) \approx V_{\mathrm{ext}}(\bm{R}_{CM},\phi,\theta,\phi)$.

\begin{figure}
\includegraphics[width=.95\columnwidth]{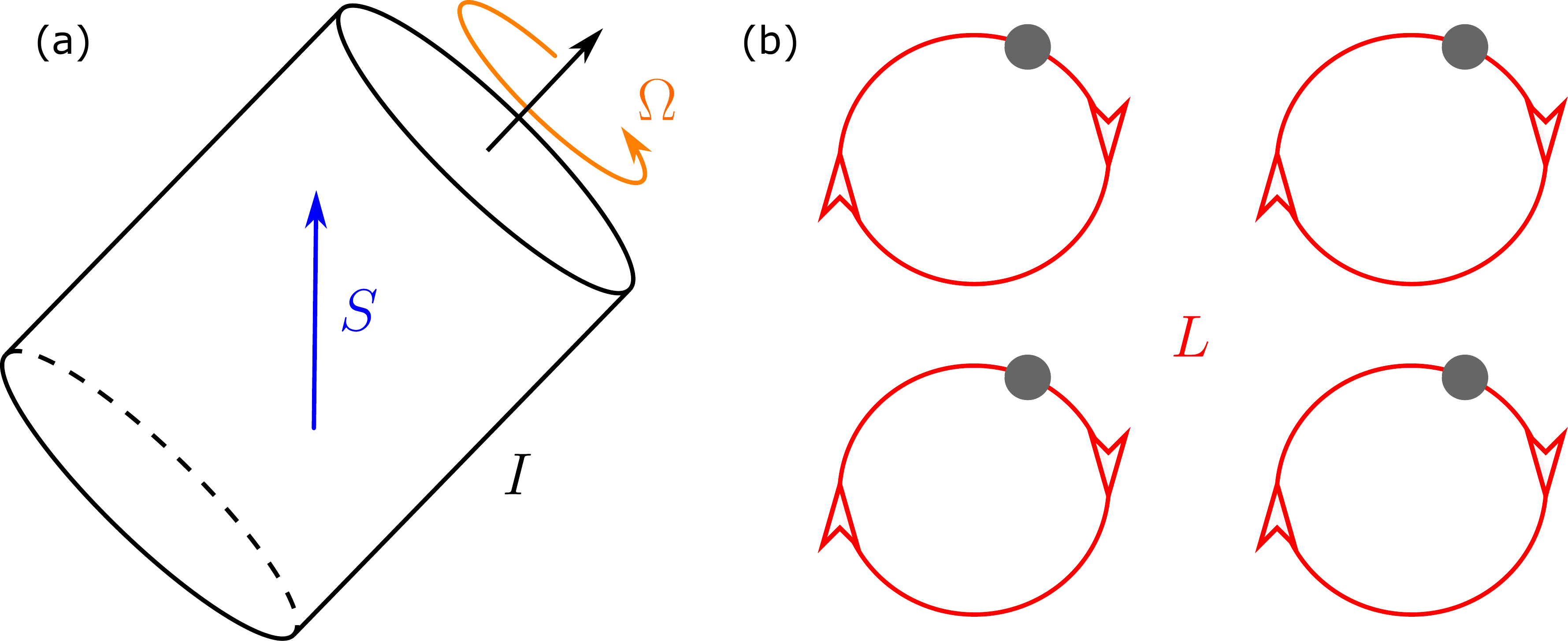}
\caption{\label{fig:Angular}
Illustration of the different kinds of angular momenta 
that are relevant for the angular momentum balance (\ref{eq:Omega}) of a magnetic insulator: 
(a) Rigid rotation with angular velocity $\bm{\Omega}$ of a cylinder 
around its axis with moment of inertia $I$ and total spin $\bm{S}$, 
(b) sketch of a phonon mode with angular momentum $\bm{L}$, showing the motion of four different volume elements 
without a global rotation as in (a). 
The total angular momentum is $\bm{J}=\mathbf{I}\cdot\bm{\Omega}+\bm{L}+\bm{S}$.}
\end{figure}

Since we describe the phonons within a rotating reference frame, it is
advantageous to also rotate the spin operators globally by the unitary
transformation:
\begin{equation} \label{eq:U}
U(\phi,\theta,\chi) = 
e^{ - \frac{i}{\hbar} \phi \sum_{j=1}^N S_j^z }
e^{ - \frac{i}{\hbar} \theta \sum_{j=1}^N S_j^y }
e^{ - \frac{i}{\hbar} \chi \sum_{j=1}^N S_j^x  } ,
\end{equation}
so that
$U^\dagger(\phi,\theta,\chi) \bm{S}_i U(\phi,\theta,\chi) = {\cal R}(\phi,\theta,\chi) \bm{S}_i$ .
As a result,
(\ref{eq:r_expansion}) and (\ref{eq:U}) transform 
$\bm{r}_{i}$ into $\bm{R}_{i}+\sum_{n=1}^{3N-6} \frac{\bm{f}_{n}(\bm{R}_{i})}{\sqrt{m_{i}}}q_{n}$ 
and $\bm{B}$ into $\mathcal{R}^{T}(\phi,\theta,\chi)\bm{B}$ in the spin Hamiltonian (\ref{eq:H_S}) 
and change the lattice kinetic energy to \cite{Watson1968,Louck1976,Littlejohn1997}
\begin{align} 
\sum_{i=1}^N \frac{\bm{p}_i^2}{2m_i} \to 
& \frac{ \bm{P}_{CM}^2 }{2M} 
+ \frac{1}{2} \bm{\Omega} \cdot \mathbf{I} \cdot \bm{\Omega}
%\left( \bm{J}' - \bm{L}' - \bm{S} \right) \cdot I^{-1} \cdot \left( \bm{J}' - \bm{L}' - \bm{S} \right)
%\nonumber\\
%
%&
+ \frac{1}{2} \sum_{n=1}^{3N-6} p_n^2
+ {\cal O}(I^{-2}) .
\label{eq:Tprime}
\end{align}
Here, $\bm{P}_{CM}=-i\hbar\partial/\partial\bm{R}_{CM}$ and $p_{n}%
=-i\hbar\partial/\partial q_{n}$ are the momentum operators of center-of-mass
translation and phonons respectively, $M=\sum_{i=1}^{N}m_{i}$ is the total
mass, and $\mathbf{I}$ is the equilibrium moment of inertia tensor
$I^{\alpha\beta}=\sum_{i=1}^{N}m_{i}\left(  \delta^{\alpha\beta}\bm{R}_{i}%
^{2}-R_{i}^{\alpha}R_{i}^{\beta}\right)  $. The latter is defined in the frame
attached to and rotating with the body, referred to as \textquotedblleft
rotating\textquotedblright, \textquotedblleft molecular\textquotedblright, or
\textquotedblleft body-fixed\textquotedblright\ frame. $\mathcal{O}(I^{-2})$
denotes correction terms originating from instantaneous phonon corrections to
the moment of inertia and quantum-mechanical commutators generated by the
nonlinear coordinate transformation (\ref{eq:r_expansion})
\cite{Watson1968,Louck1976,Littlejohn1997}. Finally,
\begin{equation} \label{eq:Omega}
\bm{\Omega}=\mathbf{I}^{-1} \cdot \left( \bm{J} - \bm{L} - \bm{S} \right)	
\end{equation}
is the vector of the angular velocity of the rigid rotation in the body-fixed
reference frame. Here the total and phonon angular momentum operators $\bm{J}$
and $\bm{L}$, and the total spin operator $\bm{S}=\sum_{i=1}^{N}\bm{S}_{i}$
are also in the body-fixed coordinate system. The three angular momenta in a
magnetic insulator are sketched in Fig.~\ref{fig:Angular}.
The total angular momentum in the laboratory frame
$\bm{J}_{\mathrm{lab}}=\mathcal{R}(\phi,\theta,\chi)\bm{J}$ obeys the standard
angular momentum algebra $[J_{\mathrm{lab}}^{x},J_{\mathrm{lab}}^{y}]=i\hbar
J_{\mathrm{lab}}^{z}$ and is conserved in the absence of external torques,
i.e., for $\bm{B}=0$ and $[\bm{J},V_{\mathrm{ext}}]=0$. The total angular
momentum in the body-fixed frame depends only on the Euler angles, i.e., it is
carried solely by the rigid-body rotation \cite{Louck1976} and the angular
momentum commutation relations are anomalous (with negative sign)
$[J^{x},J^{y}]=-i\hbar J^{z}$ \cite{VanVleck1951,Watson1968,Louck1976,Littlejohn1997}. 
Explicit expressions for the total angular momentum operators in the body-fixed and laboratory
frame are relegated to Appendix \ref{app:J}. The \textquotedblleft phonon
spin\textquotedblright\ is the phonon angular momentum in the body-fixed
frame:
\begin{equation} \label{eq:L}
\bm{L} = \sum_{i=1}^{N}\bm{l}_{i}= \sum_{i=1}^N \bm{u}_i \times \bm{\pi}_i ,
\end{equation}
where $\bm{u}_{i}$ and $\bm{\pi}_{i}$ are, respectively, the displacement and
linear momentum operators:
\begin{subequations} \label{eq:def_u_pi}
\begin{align}
\bm{u}_i =& \frac{1}{\sqrt{m_i}} \sum_{n=1}^{3N-6} \bm{f}_n(\bm{R}_i) q_n , \\
\bm{\pi}_i =& \sqrt{m_i} \sum_{n=1}^{3N-6} \bm{f}_n(\bm{R}_i) p_n .
\end{align}
\end{subequations}
Care has to be exercised when interpreting the phonon operators
(\ref{eq:def_u_pi}) and (\ref{eq:L}) in a finite system. The exclusion of the
6 degrees of freedom of the rigid-body dynamics breaks the canonical
commutation relations of the phonon position, momentum, and angular momentum
operators. Corrections of $\mathcal{O}(I^{-1})$ \cite{Watson1968} are
important for nanoscale systems. For details of the derivation of the kinetic
energy (\ref{eq:Tprime}) and the finite size corrections, we refer to
Refs.~\onlinecite{Watson1968,Louck1976,Littlejohn1997}. In the following, we
focus on systems large enough, i.e., $N\gg1$ as shown in
Appendix \ref{sec:Commutation}, to disregard finite size corrections to $\bm{L}$'s
thermal or quantum fluctuations and treat the phonon operators
(\ref{eq:def_u_pi}) and (\ref{eq:L}) canonically.

The equations of motion of the relevant angular momentum operators are now
\begin{subequations} \label{eq:Jdot}
\begin{align}
\partial_t \bm{S} + \frac{1}{2} \left( \bm{\Omega} \times \bm{S} - \bm{S} \times \bm{\Omega} \right) 
=& \bm{S} \times {\cal R}^T(\phi,\theta,\chi) \gamma\bm{B}  
\nonumber\\
&+\sum_{i=1}^N \left( \bm{R}_i + \bm{u}_i \right) \times \frac{\partial H_S}{\partial \bm{u}_i} ,
\\
\partial_t \bm{L} + \frac{1}{2} \left( \bm{\Omega} \times \bm{L} - \bm{L} \times \bm{\Omega} \right) 
=& -\sum_{i=1}^N \bm{u}_i \times \left( \frac{\partial V}{\partial \bm{u}_i} + \frac{\partial H_S}{\partial \bm{u}_i} \right) ,
\\
\partial_t \bm{J} + \frac{1}{2} \left( \bm{\Omega} \times \bm{J} - \bm{J} \times \bm{\Omega} \right)
=& \bm{S} \times {\cal R}^T(\phi,\theta,\chi) \gamma\bm{B} + \bm{{\cal T}}_{\rm ext} ,
\end{align}
\end{subequations}
where $\bm{{\cal T}}_{\mathrm{ext}}=-i[\bm{J},V_{\mathrm{ext}}]/\hbar$ is the
external mechanical torque that acts on the magnet in the body-fixed frame.
Thus, the angular momentum $\mathbf{I}\cdot\bm{\Omega}$ of the rigid rotation
satisfies
\begin{align} \label{eq:OmegaDot}
& \partial_t \left( \mathbf{I} \cdot \bm{\Omega} \right) 
+ \frac{1}{2} \left[ \bm{\Omega} \times \left( \mathbf{I} \cdot \bm{\Omega} \right) 
- \left( \mathbf{I} \cdot \bm{\Omega} \right) \times \bm{\Omega} \right] 
\nonumber\\
=& \bm{{\cal T}}_{\rm ext}
-\sum_{i=1}^N \left( \bm{R}_i \times \frac{\partial H_S}{\partial \bm{u}_i} 
- \bm{u}_i \times \frac{\partial V}{\partial \bm{u}_i} \right) .
\end{align}
Equations (\ref{eq:Jdot}) and (\ref{eq:OmegaDot}) constitute the microscopic
equations for the Einstein-de Haas \cite{Einstein1915} and Barnett
\cite{Barnett1915} effects in magnetic insulators. The left-hand sides are
covariant derivatives that account for the change in angular momentum in the
body-fixed frame \cite{Littlejohn1997}, whereas the right-hand sides are the
internal mechanical ($V$), spin-lattice ($H_{S}$), and external magnetic
($\bm{B}$) and mechanical ($\bm{{\cal T}}_{\mathrm{ext}}$) torques. The spins
exert a torque on the lattice by driving the rigid-body rotation and exciting
phonons. The torques on the right-hand sides depend on the microscopic phonon
and spin degrees of freedom that act as thermal baths and thereby break
time-reversal symmetry. We disregard radiative damping, so energy is conserved
and entropy cannot decrease, in contrast to conventional approaches to the
Einstein-de Haas effect that demand angular momentum conservation only and do
not include thermal baths. Hence, energy is not conserved in these approaches
and entropy can decrease.

\subsection{Derivation of the phenomenological theory of magnetoelasticity} 
\label{sec:Phenomenology}

Our general model for spin-lattice interactions can be parametrized by a small
number of magnetic and magnetoelastic constants at low energies.
%since the
%low-energy continuum limit of the Hamiltonian (\ref{eq:H_S}) in the body-fixed
%frame reduces to the conventional phenomenological theory of magnetoelasticity
%\cite{Eastman1966,Gurevich1996}.
In the long wavelength continuum limit $\bm{S}_{i}\rightarrow\bm{S}(\bm{r})/n$
and $\bm{u}_{i}\rightarrow\bm{u}(\bm{r})$, where $n$ is the number density of
magnetic moments. To lowest order in the gradients of spin and phonon
operators
\begin{subequations} \label{eq:Hme}
\begin{align}
H_{\textrm{Ex}} \approx
&
\frac{n}{\hbar^2 s^2} \int \textrm{d}^3r \sum_{\mu\nu} \left[
\frac{ J_{\mu\nu} }{2} + \sum_{\alpha\beta} A_{\mu\nu\alpha\beta} \epsilon^{\alpha\beta}(\bm{r}) + \ldots
\right]
\nonumber\\ 
& \phantom{\frac{n}{\hbar^2 s^2} \int \textrm{d}^3r \sum_{\mu\nu} }
\times \frac{ \partial \bm{S}(\bm{r}) }{ \partial r^\mu } \cdot \frac{ \partial \bm{S}(\bm{r}) }{ \partial r^\nu } ,
\\
H_{\textrm{A}} \approx
& \frac{n}{\hbar^2 s^2}\int \textrm{d}^3 r \sum_{\mu\nu} \left[
-\frac{ K_{\mu\nu} }{2} + \sum_{\alpha\beta} B_{\mu\nu\alpha\beta} \epsilon^{\alpha\beta}(\bm{r}) + \ldots
\right] 
\nonumber\\ 
& \phantom{\frac{n}{\hbar^2 s^2}\int \textrm{d}^3 r \sum_{\mu\nu}}
\times \tilde{S}^\mu(\bm{r}) \tilde{S}^\nu(\bm{r}) ,
\label{eq:H_A_approx}
\end{align}
\end{subequations}
where $s=Sn$ is the saturation spin density in units of $\hbar$,
\begin{equation}
\epsilon^{\alpha\beta}(\bm{r}) =
\frac{1}{2} \left[
\frac{ \partial u^\beta(\bm{r}) }{ \partial r^\alpha } +
\frac{ \partial u^\alpha(\bm{r}) }{ \partial r^\beta } +
\frac{ \partial \bm{u}(\bm{r}) }{ \partial r^\alpha } \cdot \frac{ \partial \bm{u}(\bm{r}) }{ \partial r^\beta }
\right] 
\end{equation}
is the elastic strain tensor, 
\begin{equation}
\tilde{S}^\mu(\bm{r}) = \left[S^\mu(\bm{r}) + \bm{S}(\bm{r}) \cdot 
\frac{ \partial \bm{u}(\bm{r}) }{ \partial r^\mu } \right]
\end{equation}
are the projections of the spin density on the elastically deformed anisotropy
axes, and the ellipses stand for higher powers of the strain tensor. Exchange,
anisotropy, and magnetoelastic constants can be expressed as moments of the
isotropic $\left(  J\right)  $ and anisotropic $\left(  K\right)  $ exchange
interactions and their spatial derivatives $J^{\prime}(R)=\partial
J(R)/\partial R$ and $K^{\prime}(R)=\partial K(R)/\partial R$:
\begin{subequations} \label{eq:constants}
\begin{align}
J_{\mu\nu} =& \frac{\hbar^2 s^2}{2 n^2} \sum_i J(R_i) R_i^\mu R_i^\nu ,
\\
K_{\mu\nu} =& \frac{\hbar^2 s^2}{n^2} \sum_i K(R_i) R_i^\mu R_i^\nu ,
\\
A_{\mu\nu\alpha\beta} =& \frac{\hbar^2 s^2}{4 n^2} \sum_i \frac{J'(R_i)}{R_i} R_i^\mu R_i^\nu R_i^\alpha R_i^\beta 
\\
B_{\mu\nu\alpha\beta} =& -\frac{\hbar^2 s^2}{2 n^2} \sum_i \frac{K'(R_i)}{R_i} R_i^\mu R_i^\nu R_i^\alpha R_i^\beta ,
\end{align}
\end{subequations}
The continuum limit (\ref{eq:Hme}) agrees with the standard, phenomenological
theory of magnetoelasticity \cite{Eastman1966}. Equation (\ref{eq:H_A_approx})
includes the spin-lattice coupling by rotational strains \cite{Garanin2015}
via the spin density projections $\tilde{S}^{\mu}(\bm{r})$.

The exchange, anisotropy, and magnetoelastic constants (\ref{eq:constants})
reflect the microscopic crystal symmetries. For a simple cubic lattice with
lattice constant $a=n^{-1/3}$, and nearest-neighbor isotropic as well as
next-nearest neighbor anisotropic exchange we find $J_{\mu\nu}=SJ_{s}%
\delta_{\mu\nu}$ and $K_{\mu\nu}=K\delta_{\mu\nu}$, with spin stiffness
$J_{s}=\hbar^{2}SJ(a)a^{2}$ and anisotropy constant $K=2\hbar^{2}S^{2}\left[
K(a)+4K(\sqrt{2}a)\right]  a^{2}$. The latter may be disregarded because it
only adds a constant to the Hamiltonian. The magnetoelastic coupling constants
become $A_{\mu\nu\alpha\beta}=A_{\parallel}\delta_{\mu\nu}\delta_{\nu\alpha
}\delta_{\alpha\beta}$, and $B_{\mu\nu\alpha\beta}= \left(  B_{\parallel
}-\frac{3}{2}B_{\bot}\right)  \delta_{\mu\nu} \delta_{\nu\alpha}
\delta_{\alpha\beta} +\frac{1}{2}B_{\bot}(\delta_{\mu\nu}\delta_{\alpha\beta
}+\delta_{\mu\alpha}\delta_{\nu\beta}+\delta_{\mu\beta}\delta_{\nu\alpha})$,
with $A_{\parallel}=\frac{\hbar^{2}S^{2}}{2}J^{\prime}(a)$, $B_{\parallel
}=-\hbar^{2}S^{2}\left[  K^{\prime}(a)+\sqrt{2}K^{\prime}(\sqrt{2}a)\right]
a^{3}$, and $B_{\bot}=-2\sqrt{2}\hbar^{2}S^{2}K^{\prime}(\sqrt{2}a)a^{3}$. The
anisotropy parameters $B_{\parallel}$ and $B_{\bot}$ are known for many
magnets \cite{Eastman1966}. The exchange-induced magnetoelastic constant can
be estimated as $A_{\parallel}\approx\frac{3}{2}\Gamma_{m}SJ_{s}$
\cite{Streib2019}, where $\Gamma_{m} = \partial\ln T_{C}/\partial\ln V$ is the
magnetic Gr\"{u}neisen parameter that quantifies the change of Curie
temperature $T_{C}$ with the volume $V$.

\section{Thermal spin transfer}
\label{sec:T}

\begin{figure}
\includegraphics[width=.5\columnwidth]{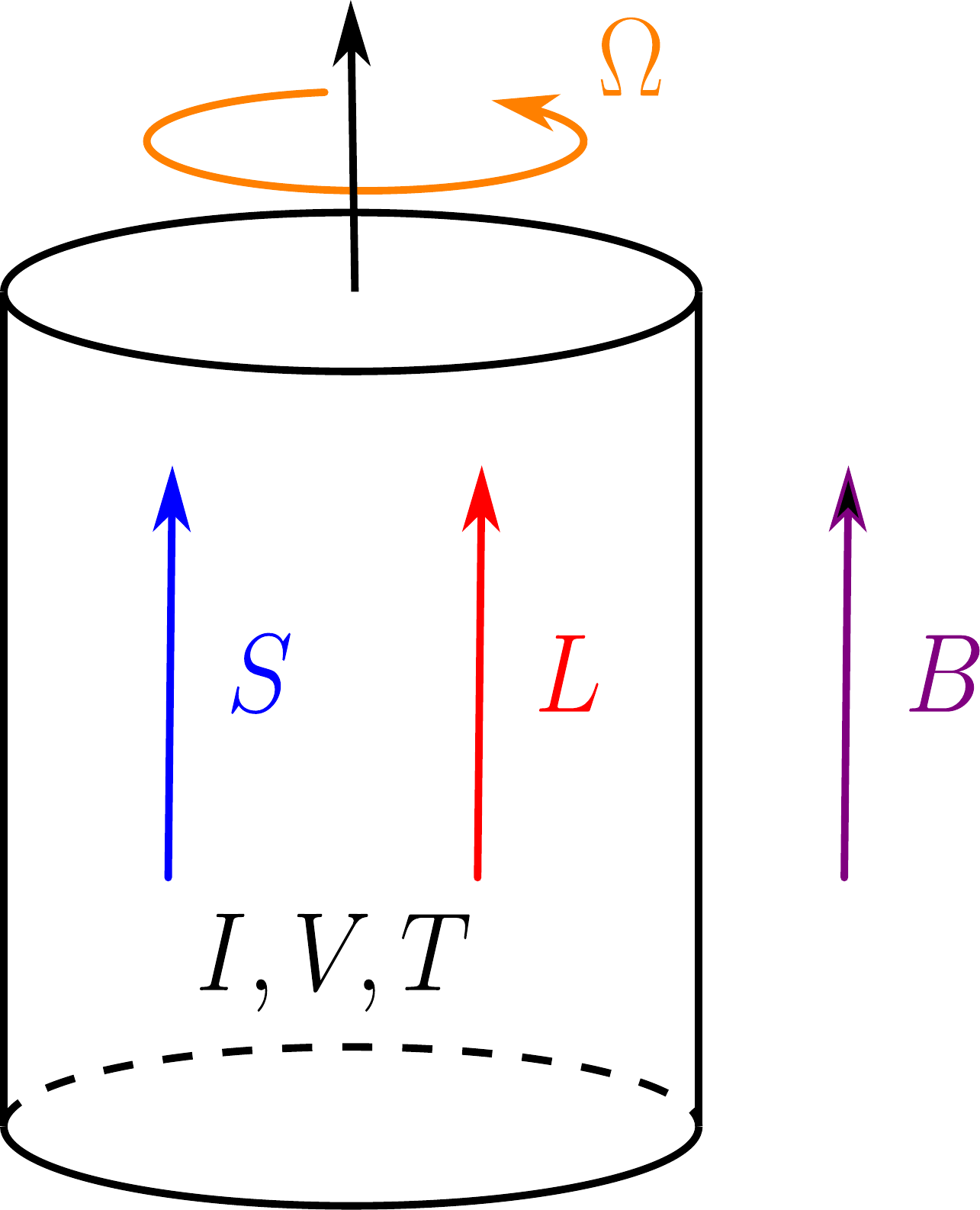}
\caption{\label{fig:Setup}
The
system under consideration in Sec.~\ref{sec:T}: A macroscopic ferromagnet with
moment of inertia $I$ and volume $V$ at temperature $T$. The total spin
$\bm{S}$ of the ferromagnet is aligned parallel to an external magnetic field
$\bm{B}$. In addition, the ferromagnet may rotate with angular velocity
$\bm{\Omega}$, and supports a phonon spin $\bm{L}$. The system is assumed to
be at rest with $\bm{\Omega}=0$ initially. Finite $\bm{\Omega}$ and $\bm{L}$
are induced by exciting the system into a nonequilibrium state at time $t=0$,
e.g., by heating the phonons or by pumping the magnons with an rf field. In
this case both $\bm{\Omega}$ and $\bm{L}$ are parallel to the total spin
$\bm{S}$ because of the conservation of total angular momentum.}
\end{figure}

In the remainder of this paper, we focus on a particular application of the
general theory, viz. the angular momentum transfer by thermal fluctuations in
the bulk of a macroscopic, externally excited, levitated ferromagnetic
particle that does not rotate ($\left\langle \bm{\Omega}\right\rangle =0$)
initially. We assume a simple cubic lattice at low temperatures. The average
magnetic order parameter, i.e., the total spin $\bm{S}$, is aligned to an
external magnetic field $\bm{B}=B\bm{\hat{e}}_{z}.$ For convenience we chose
an axially symmetric setup as sketched in Fig.~\ref{fig:Setup}.
Local spin fluctuations are described via the leading order
Holstein-Primakoff transformation \cite{Holstein1940}
\begin{subequations}
\begin{align}
S_i^+ &= (S_i^-)^\dagger = \hbar\sqrt{2S} \left[ b_i + {\cal O}(1/S) \right], \\
S_i^z &= \hbar \left(S - b_i^\dagger b_i \right) ,
\end{align}
\end{subequations}
where $b_{i}^{\dagger}\left(  b_{i}\right)  $ is the magnon creation
(annihilation) operator on site $i$, which satisfies the Boson commutation
relations $[b_{i},b_{j}^{\dagger}]=\delta_{ij}$.

In a macroscopic magnet the time scales between the rigid-body rotation and
the internal magnon and phonon dynamics are decoupled: For a system with
volume $V$, the moment of inertia $I\sim V^{5/3}$, whereas the angular
momentum operators $\bm{J}$, $\bm{L}$, and $\bm{S}$ are extensive quantities,
proportional to $V$. According to Eq.~(\ref{eq:Omega}) the angular velocity
$\bm{\Omega}$ scales as $V^{-2/3}$. On the other hand, the lowest phonon
frequency $\omega_{\mathrm{min}}\sim V^{-1/3}$, while the magnon gap is
controlled by external magnetic and internal anisotropy fields and is
typically of the order of $10\,\text{GHz}$ independent of $V$. For
sufficiently large systems and weak driving, inertial forces of the rigid-body
rotation therefore affect the dynamics of both magnons and phonons only
negligibly and can be disregarded. By the same argument, the energy $\frac
{1}{2}\bm{\Omega}\cdot\mathbf{I}\cdot\bm{\Omega}\sim V^{1/3}$ of the
rigid-body rotation is small compared to the total magnon and phonon energies
$\sim V$. Energy is then (almost) exclusively equilibrated by spin-phonon
interactions, under the constraint of angular momentum conservation that
includes the rigid-body rotation. For example, consider the change in energy
of the magnet at rest when a single magnon with frequency $\epsilon$ is
removed from the system, which increases the spin by $\Delta S=\hbar.$ If this
angular momentum is fully transferred to the rigid rotation of a sphere with
scalar moment of inertia $I$, $\Delta L_{R}=-\hbar=I\Omega^{z}$. For a
macroscopic magnet the change of rotational energy $\Delta E_{R}=\hbar^{2}/2I$
is negligible compared to the magnetic energy change $\Delta E_{m}%
=-\hbar\epsilon$, since the typical magnon frequencies are in the $\text{GHz}%
$-$\text{THz}$ range, whereas both $\Omega^{z}$ and $I^{-1}$ are small by some
power of the inverse volume. Consequently, the energy of the magnon cannot be
transferred completely to the rigid rotation, since both energy and angular
momentum cannot be conserved simultaneously. The Einstein-de Haas effect can
therefore not exist without an intermediate bath, which in magnetic insulators
can only be the lattice vibrations.

At temperatures sufficiently below the Curie and Debye temperatures and weak
external excitation, only the long-wavelength modes are occupied and
Eq.~(\ref{eq:Hme}) is appropriate. At not too low temperatures we may also
disregard magnetodipolar interactions \cite{Streib2019}. We assume again that
the magnet is sufficiently large that surface effects are small and the
eigenmodes of the system may be approximated by plane waves. Then the Fourier
transform $b_{i}=N^{-1/2}\sum_{\bm{k}}e^{-i\bm{k}\cdot\bm{R}_{i}}b_{\bm{k}}$
leads to the magnetic Hamiltonian:
\begin{equation}
H_m = \sum_{\bm{k}} \hbar \epsilon_{\bm{k}} b_{\bm{k}}^\dagger b_{\bm{k}} ,
\end{equation}
where $\epsilon_{\bm{k}}=\gamma B+J_{s}\bm{k}^{2}/\hbar$ is the magnon
frequency dispersion relation and $b_{\bm{k}}^{\dagger}\left(  b_{\bm{k}}%
\right) $ are creation (annihilation) operators of a magnon with wave vector
$\bm{k}$.

Analogously, the finite size of a sufficiently large system only affects
phonons with wavelengths ${\cal O}\left(  V^{1/3}\right)  $ and a small density of
states. We may then expand $\bm{u}_{i}=N^{-1/2}\sum_{\bm{k}}e^{-i\bm{k}\cdot
\bm{R}_{i}}\bm{u}_{\bm{k}}$, with
\begin{equation}
\bm{u}_{\bm{k}} = \sum_\lambda \sqrt{\frac{\hbar}{2m\omega_{\bm{k}\lambda}}} \bm{\hat{e}}_{\bm{k}\lambda}
\left( a_{\bm{k}\lambda} + a^\dagger_{-\bm{k}\lambda} \right) .
\end{equation}
Here, $a_{\bm{k}\lambda}^{\dagger}\left(  a_{\bm{k}\lambda}\right)  $ creates
(annihilates) a phonon with momentum $\bm{k}$, polarization vector
$\bm{\hat{e}}_{\bm{k}\lambda}$, and frequency $\omega_{\bm{k}\lambda}$, and
Bose commutation relations $[a_{\bm{k}\lambda},a_{\bm{k}^{\prime}%
\lambda^{\prime}}^{\dagger}]= \delta_{\bm{k}\bm{k}^{\prime}}\delta
_{\lambda\lambda^{\prime}}$. An isotropic elastic solid supports three
acoustic phonon branches: two degenerate transverse ($\lambda=\pm$) and one
longitudinal ($\lambda=\parallel$) one, with $\omega_{\bm{k}\lambda
}=c_{\lambda}k$, where the $c_{\lambda}$ are the sound velocities. We choose a
circular basis $\lambda=\pm$ for the transverse phonons
\cite{Levine1962,McLellan1988,Garanin2015} and express the momentum $\bm{k}$
in spherical coordinates, so that
\begin{subequations} \label{eq:pol}
\begin{align}
\bm{\hat{e}}_{\bm{k}\pm} =& \frac{1}{\sqrt{2}} \bigl[
\bm{\hat{e}}_x \left( \cos\theta_{\bm{k}}\cos\phi_{\bm{k}} \mp i \sin\phi_{\bm{k}} \right) 
\nonumber\\
&
\phantom{\frac{1}{\sqrt{2}} \bigl[} +
\bm{\hat{e}}_y \left( \cos\theta_{\bm{k}}\sin\phi_{\bm{k}} \pm i \cos\phi_{\bm{k}} \right) -
\bm{\hat{e}}_z \sin\theta_{\bm{k}}
\bigr] ,
\\
\bm{\hat{e}}_{\bm{k}\parallel} =& i \left[
\bm{\hat{e}}_x \sin\theta_{\bm{k}}\cos\phi_{\bm{k}} +
\bm{\hat{e}}_y \sin\theta_{\bm{k}}\sin\phi_{\bm{k}} +
\bm{\hat{e}}_z \cos\theta_{\bm{k}} 
\right] 
\nonumber\\
=& i \frac{\bm{k}}{k} .
\end{align}
\end{subequations}
In this basis the phonon spin (\ref{eq:L}) is diagonal \cite{Levine1962,McLellan1988,Garanin2015}:
\begin{equation} \label{eq:L_Bulk}
\bm{L} = - \hbar \sum_{\bm{k}} \frac{\bm{k}}{k} 
\left( a^\dagger_{\bm{k}+} a_{\bm{k}+} - a^\dagger_{\bm{k}-} a_{\bm{k}-} \right) ,
\end{equation}
where we dropped terms that have vanishing expectation values for
noninteracting phonons. Analogous to photons, circularly polarized phonons
with $\lambda=\pm$ carry one spin quantum $\mp\hbar$ parallel to their wave
vector that is carried exclusively by transverse phonons.
Mentink \textit{et al.} \cite{Mentink2019} report that only longitudinal phonons contribute to the electron-phonon spin transfer. 
This is not a contradiction, because they define phonon angular momentum different from Eq.~(\ref{eq:L}) as adhered to in most papers \cite{Garanin2015,Levine1962,McLellan1988}. 
On the other hand, that definition appears similar to the field (or pseudo) angular momentum introduced independently by Nakane and Kohno \cite{Nakane2018}.
%Mentink \textit{et al.}'s \cite{Mentink2019} claim that only longitudinal phonons contribute to the electron-phonon spin transfer is %rooted in a definition of the phonon angular momentum that differs from the standard (\ref{eq:L}) %\cite{Garanin2015,Levine1962,McLellan1988}, 
%and is similar to the field (or pseudo) angular momentum introduced independently by Nakane and Kohno \cite{Nakane2018}.

\begin{figure}
\includegraphics[width=.75\columnwidth]{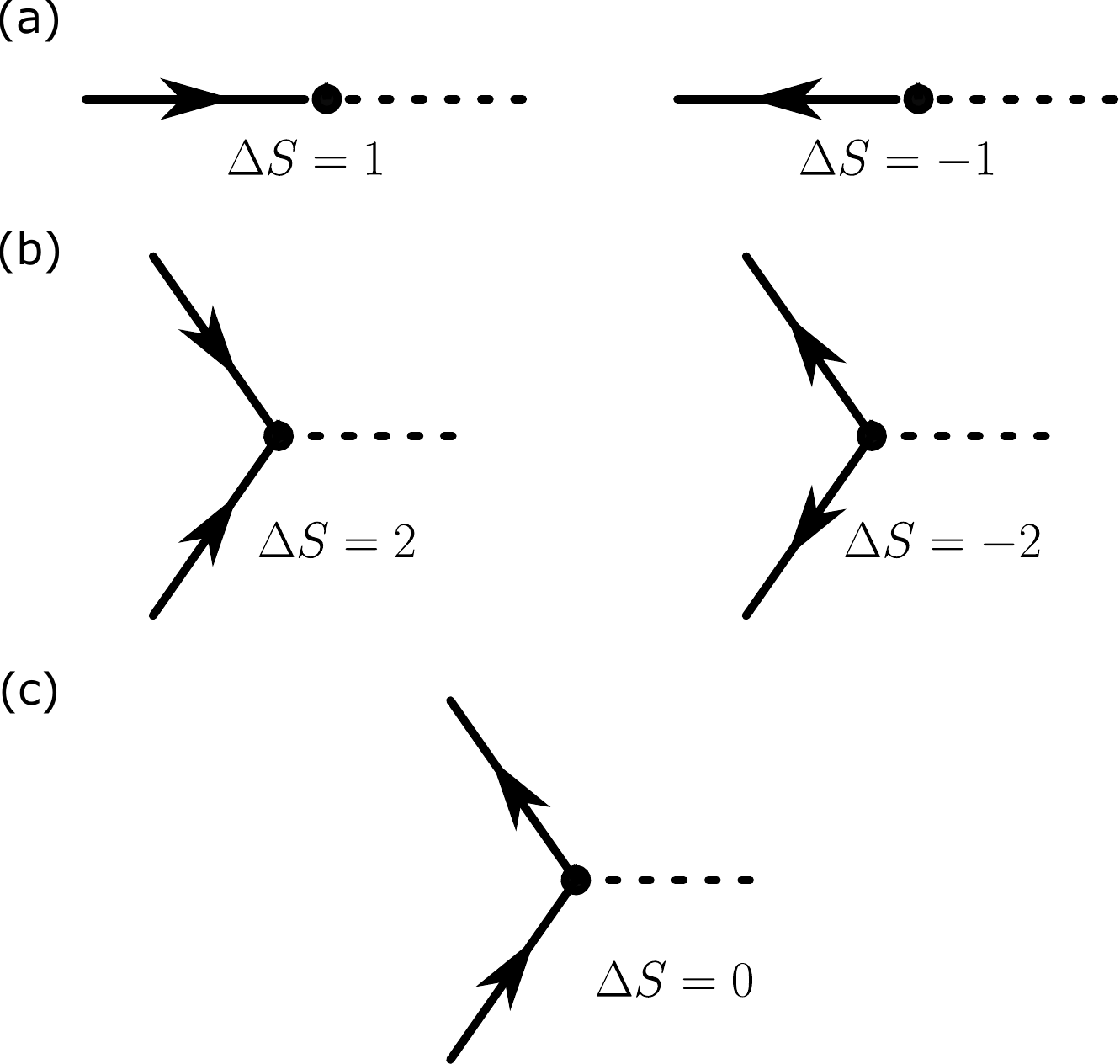}
\caption{\label{fig:Vertices}
Diagrams
of the leading magnon-phonon scattering processes in Eq. (\ref{eq:Hmp}), 
with corresponding change of electron spin $\Delta S$.
Solid, directed lines denote magnons and dashed lines phonons.
(a) Magnon-phonon hybridization,
(b) magnon-number nonconserving confluence processes, and
(c) magnon-number conserving Cherenkov scattering.}
\end{figure}

The leading one-phonon/one- and two-magnon contributions to the magnetoelastic
Hamiltonian (\ref{eq:Hme}) read in momentum space
\begin{align}
H_{mp} =
& 
\sum_{\bm{k}} \left( \bm{\Gamma}_{\bm{k}} b_{\bm{k}}
+ \bm{\Gamma}_{-\bm{k}}^* b_{-\bm{k}}^\dagger \right) \cdot \bm{u}_{-\bm{k}} 
\nonumber\\
&
+ \frac{1}{\sqrt{N}} \sum_{\bm{k}\bm{k}'} \biggl(
\bm{U}_{\bm{k},\bm{k}'} \cdot \bm{u}_{\bm{k}-\bm{k}'} b_{\bm{k}}^\dagger \hat{b}_{\bm{k}'} 
\nonumber\\
&
+ \frac{1}{2} \bm{V}_{\bm{k},\bm{k}'} \cdot \bm{u}_{-\bm{k}-\bm{k}'} b_{\bm{k}} b_{\bm{k}'}
+ \frac{1}{2} \bm{V}_{\bm{k},\bm{k}'}^* \cdot \bm{u}_{\bm{k}+\bm{k}'} b_{\bm{k}}^\dagger b_{\bm{k}'}^\dagger
\biggr) ,
\label{eq:Hmp}
\end{align}
with interaction vertices 
\begin{subequations}
\begin{align}
\bm{\Gamma}_{\bm{k}} =&
- \frac{i B_\bot}{\sqrt{2S}} \left[ 
\left( \bm{\hat{e}}_x - i \bm{\hat{e}}_y \right) k^z + \bm{\hat{e}}_z \left( k^x - i k^y \right) \right] ,
\\
\bm{U}_{\bm{k}+\bm{q},\bm{k}} =&
\frac{ i B_\parallel }{S} \left( \bm{\hat{e}}_x q^x + \bm{\hat{e}}_y q^y - 2 \bm{\hat{e}}_z q^z \right)
\nonumber\\
&
+ \frac{2i A_\parallel}{S} \sum_\alpha \bm{\hat{e}}_\alpha \left( k^\alpha + q^\alpha \right) k^\alpha q^\alpha ,
\\
\bm{V}_{\bm{k}+\bm{q},-\bm{k}} =&
- \frac{ i B_\parallel}{S} \left( \bm{\hat{e}}_x q^x - \bm{\hat{e}}_y q^y \right)
- \frac{ B_\bot}{S} \left( \bm{\hat{e}}_x q^y + \bm{\hat{e}}_y q^x \right) .
\end{align}
\end{subequations}
The first line of the magnetoelastic Hamiltonian (\ref{eq:Hmp}) describes the
hybridization of magnons and phonons or magnon polaron \cite{Kikkawa2016},
while the second and third line are, respectively, magnon-number conserving
Cherenkov scattering and magnon-number nonconserving confluence processes
\cite{Gurevich1996} as illustrated by Fig.~\ref{fig:Vertices}. We disregard
the weak two-phonon one-magnon scattering processes \cite{Streib2019}.
Angular momentum is transferred between the magnetic order and the
lattice by the magnon-number nonconserving hybridization and confluence
processes, while magnon-number conserving scatterings transfer energy only.
Energy conservation requires that phonons have frequencies larger than $\gamma
B$ ($2\gamma B$), i.e., in the GHz range, in order to participate resonantly
(by confluence) to the spin transfer. The applied magnetic field is an
important control parameter since above the critical value
\begin{equation}
B_{c,\lambda} = \frac{ \hbar c_\lambda^2 }{ 4 \gamma J_s }
\end{equation}
hybridization and confluence processes are forbidden for phonons with
polarization $\lambda$. For $B_{\mathrm{ext}}>B_{c,\bot},B_{c,\parallel}$
other spin transfer mechanisms must be invoked, such as interface/surface
\cite{Streib2018} or higher order magnon-phonon scattering \cite{Streib2019}.

\subsection{Kinetic equations}

Treating the magnon-phonon interaction Hamiltonian (\ref{eq:Hmp}) by Fermi's
golden rule leads to rate equations for the bulk magnon and phonon
distribution functions $n_{\bm{k}}=\left\langle b_{\bm{k}}^{\dagger}%
b_{\bm{k}}\right\rangle $ and $N_{\bm{k}\lambda}=\left\langle a_{\bm{k}\lambda
}^{\dagger}a_{\bm{k}\lambda}\right\rangle $ \cite{Gurevich1996}:
\begin{align}
\dot{n}_{\bm{k}} =
&
\sum_\lambda \frac{ \pi | \bm{\hat{e}}_{\bm{k}\lambda}^* \cdot \bm{\Gamma}_{\bm{k}} |^2 }
{ m \hbar \omega_{\bm{k}\lambda} } 
\delta\left( \epsilon_{\bm{k}} - \omega_{\bm{k}\lambda} \right) \left( N_{\bm{k}\lambda} - n_{\bm{k}} \right) 
\nonumber\\
&
+
\frac{1}{N} \sum_{\bm{k}'\bm{q}\lambda} \delta_{\bm{k}-\bm{k}',\bm{q}}
\frac{ \pi | \bm{\hat{e}}_{\bm{q}\lambda} \cdot \bm{U}_{\bm{k},\bm{k}'} |^2 }{ m \hbar \omega_{\bm{q}\lambda} }
\delta\left( \epsilon_{\bm{k}} - \epsilon_{\bm{k}'} - \omega_{\bm{q}\lambda} \right)
\nonumber\\
&
\phantom{+}\times
\left[ \left( 1 + n_{\bm{k}} \right) n_{\bm{k}'} N_{\bm{q}\lambda} - 
n_{\bm{k}} \left( 1 + n_{\bm{k}'} \right) \left( 1 + N_{\bm{q}\lambda} \right) \right]
\nonumber\\
&
+
\frac{1}{N} \sum_{\bm{k}'\bm{q}\lambda} \delta_{\bm{k}-\bm{k}',\bm{q}}
\frac{ \pi | \bm{\hat{e}}_{\bm{q}\lambda} \cdot \bm{U}_{\bm{k},\bm{k}'} |^2 }{ m \hbar \omega_{\bm{q}\lambda} }
\delta\left( \epsilon_{\bm{k}} - \epsilon_{\bm{k}'} + \omega_{\bm{q}\lambda} \right)
\nonumber\\
&
\phantom{+}\times
\left[ \left( 1 + n_{\bm{k}} \right) n_{\bm{k}'} \left( 1 + N_{-\bm{q}\lambda} \right) - 
n_{\bm{k}} \left( 1 + n_{\bm{k}'} \right) N_{-\bm{q}\lambda} \right] 
\nonumber\\
&
+
\frac{1}{N} \sum_{\bm{k}'\bm{q}\lambda} \delta_{\bm{k}+\bm{k}',\bm{q}}
\frac{ \pi | \bm{\hat{e}}_{\bm{q}\lambda}^* \cdot \bm{V}_{\bm{k},\bm{k}'} |^2 }{ m \hbar \omega_{\bm{q}\lambda} }
\delta\left( \epsilon_{\bm{k}} + \epsilon_{\bm{k}'} - \omega_{\bm{q}\lambda} \right)
\nonumber\\
&
\phantom{+}\times
\left[ \left( 1 + n_{\bm{k}} \right) \left( 1 + n_{\bm{k}'} \right) N_{\bm{q}\lambda} - 
n_{\bm{k}} n_{\bm{k}'} \left( 1 + N_{\bm{q}\lambda} \right) \right] ,
\label{eq:ndot}
\end{align}
and
\begin{align}
\dot{N}_{\bm{q}\lambda} =
&
\frac{ \pi | \bm{\hat{e}}_{\bm{q}\lambda}^* \cdot \bm{\Gamma}_{\bm{q}} |^2 }{ m \hbar \omega_{\bm{q}\lambda} }
\delta\left( \epsilon_{\bm{q}} - \omega_{\bm{q}\lambda} \right) \left( n_{\bm{q}} - N_{\bm{q}\lambda} \right) 
\nonumber\\
&
+
\frac{1}{N} \sum_{\bm{k}\bm{k}'} \delta_{\bm{k}-\bm{k}',\bm{q}} 
\frac{ \pi | \bm{\hat{e}}_{\bm{q}\lambda} \cdot \bm{U}_{\bm{k},\bm{k}'} |^2 }{ m \hbar \omega_{\bm{q}\lambda} }
\delta\left( \epsilon_{\bm{k}} - \epsilon_{\bm{k}'} - \omega_{\bm{q}\lambda} \right)
\nonumber\\
&
\phantom{+}\times
\left[ n_{\bm{k}} \left( 1 + n_{\bm{k}'} \right) \left( 1 + N_{\bm{q}\lambda} \right) -
\left( 1 + n_{\bm{k}} \right) n_{\bm{k}'} N_{\bm{q}\lambda} \right]
\nonumber\\
&
+
\frac{1}{N} \sum_{\bm{k}\bm{k}'} \delta_{\bm{k}+\bm{k}',\bm{q}} 
\frac{ \pi | \bm{\hat{e}}_{\bm{q}\lambda}^* \cdot \bm{V}_{\bm{k},\bm{k}'} |^2 }{ 2 m \hbar \omega_{\bm{q}\lambda} }
\delta\left( \epsilon_{\bm{k}} + \epsilon_{\bm{k}'} - \omega_{\bm{q}\lambda} \right)
\nonumber\\
&
\phantom{+}\times
\left[ n_{\bm{k}} n_{\bm{k}'} \left( 1 + N_{\bm{q}\lambda} \right) - 
\left( 1 + n_{\bm{k}} \right) \left( 1 + n_{\bm{k}'} \right) N_{\bm{q}\lambda} \right] .
\label{eq:Ndot}
\end{align}
The first term on the right-hand side of both Eqs.~(\ref{eq:ndot}) and
(\ref{eq:Ndot}) is caused by the direct magnon-phonon conversion process in
Fig.~\ref{fig:Vertices}(a) that gives rise to magnetoelastic waves (magnon
polarons) \cite{Ruckriegel2014,Kikkawa2016}. It diverges because perturbation
theory breaks down at the crossing of magnon and phonon modes. The singularity
can be removed by choosing a basis that diagonalizes the Hamiltonian
\cite{Flebus2017}. Here we regularize it with a finite broadening
\cite{Schmidt2018} that is larger than the magnon-polaron gap, which leads to
well-behaved integrated quantities such as energy, momentum, and spin densities.

\subsection{Linear response}

We capture the dynamics of energy and spin relaxation in linear response to
weak perturbations, assuming that magnons and phonons stay close to a common
thermal equilibrium at temperature $T$. The spin-lattice interaction transfers
both energy and angular momentum which changes magnon and phonon energy
\begin{subequations}
\begin{align}
\delta E_m(t) =& \frac{1}{V} \sum_{\bm{k}} \hbar \epsilon_{\bm{k}} 
\left[ n_{\bm{k}}(t) - f_B\left( \frac{ \hbar \epsilon_{\bm{k}} }{ k_B T } \right) \right], 
\label{eq:E_m}
\\
\delta E_\lambda(t) =& \frac{1}{V} \sum_{\bm{k}} \hbar \omega_{\bm{k}\lambda} 
\left[ N_{\bm{k}\lambda}(t) - f_B\left( \frac{ \hbar \omega_{\bm{k}\lambda} }{ k_B T } \right) \right], 
\label{eq:E_lambda}
\end{align}
\end{subequations}
as well as spin densities 
\begin{subequations}
\begin{align}
\delta s(t) =& \frac{\hbar}{V} \sum_{\bm{k}}  
\left[ n_{\bm{k}}(t) - f_B\left( \frac{ \hbar \epsilon_{\bm{k}} }{ k_B T } \right) \right], 
\label{eq:s}
\\
\delta l(t) =& \frac{\langle L^z (t) \rangle}{V} = -\frac{\hbar}{V} \sum_{\bm{k}} \frac{k^z}{k} 
\left[ N_{\bm{k}+}(t) - N_{\bm{k}-}(t) \right] , 
\label{eq:Ldensity}
\end{align}
\end{subequations}
where $f_{B}(x)=1/\left(  e^{x}-1\right)  $ is the Bose distribution function.
Since a precessing magnetic moment can inject a transverse, circularly
polarized momentum current into an adjacent nonmagnetic insulator
\cite{Streib2018} we also consider transverse phonon momentum densities
\begin{equation} \label{eq:rho_pm}
\rho_\pm(t) = 
\frac{1}{V} \sum_{\bm{k}} \hbar k^z N_{\bm{k}\pm}(t) ,
\end{equation}
with $\rho_{+}=-\rho_{-}\ $at equilibrium.
In a driven system $\delta\rho
=\rho_{+}-\rho_{-}$ can be finite as can be seen from the interaction vertex
between magnons and transverse phonons in equations (\ref{eq:ndot}) and (\ref{eq:Ndot}):
\begin{equation}
\left| \bm{\hat{e}}_{\bm{k}\pm}^* \cdot \bm{\Gamma} \right|^2 = 
\frac{ B_\bot^2 k^2}{4 S} \left( \cos^2\theta_{\bm{k}} - \sin^2\theta_{\bm{k}} \mp \cos\theta_{\bm{k}} \right)^2 . 
\end{equation}
For phonons propagating along $\pm\bm{\hat{e}}_{z}$, i.e., with $\cos
\theta_{\bm{k}}=\pm1$, this expression is only finite for polarization
direction $\lambda=\mp$. Hence, magnons couple to $\lambda=+$ ($\lambda=-$)
phonons traveling in the $-\bm{\hat{e}}_{z}$ ($+\bm{\hat{e}}_{z}$) direction. An
imbalance in the magnon distribution thus creates transverse phonon
polarizations $\delta\rho$ and a finite phonon spin polarization
(\ref{eq:Ldensity}).

The coupled kinetic equations (\ref{eq:ndot}) and (\ref{eq:Ndot}) can be
simplified by assuming that magnon-magnon and phonon-phonon interactions
thermalize the distributions to a quasiequilibrium form that can be
parametrized by slowly varying variables conjugate to the macroscopic
observables of interest \cite{Schmidt2018,Cornelissen2016}, i.e., the energy
and spin densities given in Eqs.~(\ref{eq:E_m}) and (\ref{eq:s}),
respectively. These conjugate variables are a temperature deviation $\delta
T_{m}(t)$ and a magnon chemical potential $\mu(t)$ \cite{Cornelissen2016},
such that
\begin{equation} \label{eq:n_ansatz}
n_{\bm{k}}(t) = f_B\left( \frac{\hbar\epsilon_{\bm{k}}-\mu(t)}{k_B (T+\delta T_m(t))} \right) .
\end{equation}
This parametrization of the magnon distribution is accurate for thermal
magnons when the number-conserving exchange interaction is the dominant
scattering mechanism, which is usually the case in magnetic insulators
\cite{Gurevich1996,Cornelissen2016}.

A parametrization such as Eq. (\ref{eq:n_ansatz}) of the phonon distribution
fails because a phonon chemical potential does not lead to a finite phonon
spin polarization $\delta\rho$ because the angular dependence, $k_{z}%
\propto\cos\theta_{\bm{k}}$ in Eq. (\ref{eq:rho_pm}), averages to zero when
the distribution $N_{\bm{k}\pm}$ is isotropic in momentum space. We therefore
focus on the leading anisotropic term, which is a Bose distribution rigidly
shifted by a polarization-dependent phonon drift velocity $v_{\lambda}$:
\begin{equation} \label{eq:N_ansatz}
N_{\bm{k}\lambda}(t) = f_B\left( \frac{\hbar\omega_{\bm{k}\lambda}-\hbar v_\lambda(t) k^z}{k_B (T+\delta T_\lambda(t))} \right) .
\end{equation}
Because the transverse phonon modes are degenerate, we set $\delta
T_{+}(t)=\delta T_{-}(t)\equiv\delta T_{\bot}(t)$ without loss of generality,
but we allow for different temperatures of longitudinal and transverse
phonons, $\delta T_{\parallel}(t)$ and $\delta T_{\bot}(t)$, and associated
energy densities. Global linear momentum conservation requires $v_{+}%
(t)=-v_{-}(t)\equiv v(t)$ and $v_{\parallel}(t)=0$. Just as for the magnon
distribution function (\ref{eq:n_ansatz}), the parametrization
(\ref{eq:N_ansatz}) of the phonon distribution function contains some tacit
assumptions about the relative importance of different scattering mechanisms:
In particular, it should be applicable when polarization- and
momentum-conserving phonon-phonon scattering dominates over the nonconserving
scattering mechanisms. In YIG, the acoustic quality is much better than the
magnetic one \cite{Kikkawa2016}, which supports our shifted-distribution
ansatz (\ref{eq:N_ansatz}). Also, a finite drift velocity $v$ implies
existence of a phonon current on relatively large time scales, which requires
a system size $\sim V^{1/3}$ larger than the phonon relaxation length.

The response to leading order in the nonequilibrium parameters reads
\begin{subequations} \label{eq:observables}
\begin{align}
\delta E_m(t) =& - \frac{1}{V} \sum_{\bm{k}} f'_B\left( \frac{\hbar\epsilon_{\bm{k}}}{k_B T} \right) \hbar \epsilon_{\bm{k}}
\nonumber\\
& \phantom{ - \frac{1}{V} \sum_{\bm{k}} } \times
\left( \frac{\hbar \epsilon_{\bm{k}}}{k_B T} \frac{\delta T_m(t)}{T} + \frac{\mu(t)}{k_B T} \right) ,
\\ 
\delta E_\lambda(t) =& - \frac{1}{V} \sum_{\bm{k}} f'_B\left( \frac{\hbar\omega_{\bm{k}\lambda}}{k_B T} \right) 
\hbar \omega_{\bm{k}\lambda}
\frac{\hbar \omega_{\bm{k}\lambda}}{k_B T} \frac{\delta T_\lambda(t)}{T} ,
\\
\delta s(t) =& - \frac{\hbar}{V} \sum_{\bm{k}} f'_B\left( \frac{\hbar\epsilon_{\bm{k}}}{k_B T} \right)
\left( \frac{\hbar \epsilon_{\bm{k}}}{k_B T} \frac{\delta T_m(t)}{T} + \frac{\mu(t)}{k_B T} \right) ,
\\
\delta \rho(t) =& - \frac{2\hbar}{V} \sum_{\bm{k}} f'_B\left( \frac{\hbar\omega_{\bm{k}\bot}}{k_B T} \right) (k^z)^2
\frac{\hbar v(t)}{k_B T} ,
\end{align}
\end{subequations}
where $f_{B}^{\prime}(x)=\partial f_{B}(x)/\partial x$, 
and the nonequilibrium phonon spin density is
\begin{equation} \label{eq:deltaL}
\delta l(t) 
= \frac{2\hbar}{V} \sum_{\bm{k}} f'_B\left( \frac{\hbar\omega_{\bm{k}\bot}}{k_B T} \right) \frac{(k^z)^2}{k}
\frac{\hbar v(t)}{k_B T} .
\end{equation}
According to Eq.~(\ref{eq:Omega}), the angular momentum of the rigid body
rotation around a principal axis of the tensor of inertia is 
$I\Omega^{z}(t)=V[\delta j_{0}+\delta s(t)-\delta l(t)]$, 
where $\delta j_{0}$ is an angular momentum density injected by external torques.

The linear response can be summarized by 
\begin{widetext}
\begin{equation} \label{eq:Linear_Response}
\partial_t 
\left( \begin{matrix}
\delta E_m(t) / k_B T \\ 
\delta E_\bot(t) / k_B T \\ 
\delta E_\parallel(t) / k_B T \\ 
\delta s(t) / \hbar \\ 
c_\bot \delta \rho(t) / k_B T
\end{matrix} \right)
=
-
\left( \begin{matrix}
\Gamma_{\bot} + \Gamma_{\parallel} & -\Gamma_{\bot} & -\Gamma_{\parallel} 
& -\Gamma_{\bot\mu} - \Gamma_{\parallel\mu} & -\Gamma_{\bot v} \\
-\Gamma_{\bot} & \Gamma_{\bot} & 0 & \Gamma_{\bot\mu} & \Gamma_{\bot v} \\
-\Gamma_{\parallel} & 0 & \Gamma_{\parallel} & \Gamma_{\parallel\mu} & 0 \\
-\Gamma_{\bot\mu} - \Gamma_{\parallel\mu} & \Gamma_{\bot\mu} & \Gamma_{\parallel\mu} & \Gamma_{\mu} & \Gamma_{v \mu} \\
-\Gamma_{\bot v} & \Gamma_{\bot v} & 0 & \Gamma_{v\mu} & \Gamma_{v} \\
\end{matrix} \right)
\left( \begin{matrix}
\delta T_m(t) / T \\ 
\delta T_\bot(t) / T \\ 
\delta T_\parallel(t) / T \\ 
\mu(t) / k_B T \\ 
v(t) / c_\bot
\end{matrix} \right) ,
\end{equation}
\end{widetext}
where $\delta E_{\bot}=\delta E_{+}+\delta E_{-}$is the change in the
transverse phonon energy density. The Onsager-reciprocal relaxation rates
$\Gamma_{\alpha\beta}$ from Eqs.~(\ref{eq:ndot}) and (\ref{eq:Ndot}) are
listed in Appendix \ref{sec:Rates}.

\begin{figure*}
\includegraphics[width=.82\textwidth]{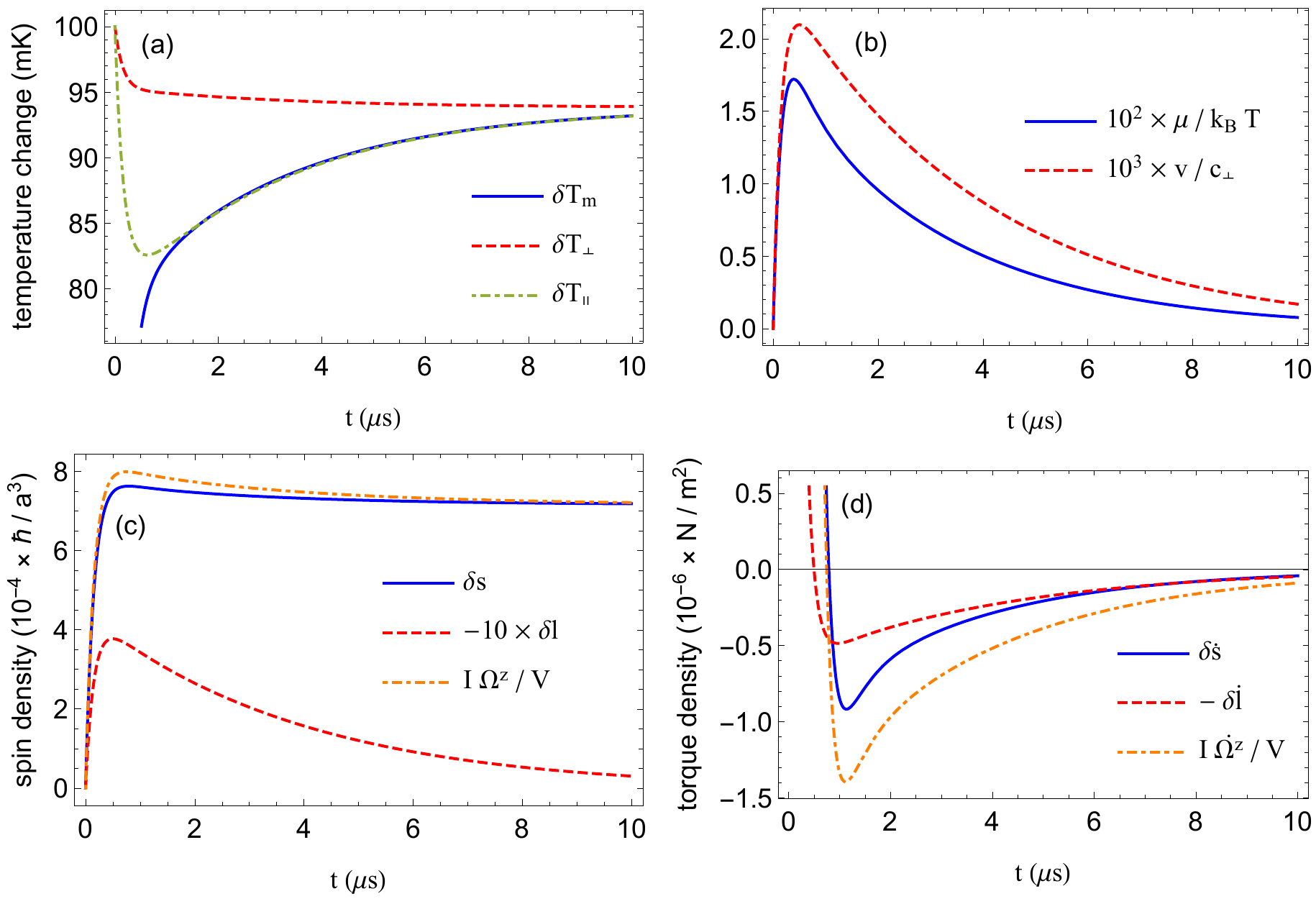}
\caption{\label{figHeating} Nonequilibrium dynamics of 
(a) magnon and phonon temperatures,
(b) magnon chemical potential and phonon drift velocity,
(c) spin densities, and 
(d) torque densities 
of magnons, phonons, and rigid rotation,
for heating initial conditions 
$\delta T_\bot(t=0) = \delta T_\parallel(t=0) = 1\,\textrm{K}$,
and $\delta T_m(t=0)=\mu(t=0)=v(t=0)=0$.
Temperature and external magnetic field are $T=10\,\textrm{K}$ and $B=1\,\textrm{T}$.}
\end{figure*}

In the following, we discuss the solutions for the material parameters of
yttrium-iron garnet in Table \ref{tab:YIG_values}.
\begin{table}
\caption{Magnetic and elastic material parameters of yttrium-iron garnet,
adopted from Refs.~\onlinecite{Gurevich1996,Cornelissen2016,Cherepanov1993,Maier-Flaig2017,Streib2019}.
If not indicated otherwise, the parameters are measured at room temperature.}
\begin{tabular}{ l c c c }
\hline\hline
																							& Symbol 						& Value 									& Unit 													\\
\hline
lattice constant 															& $a$ 							& $12.376$ 								& $\text{\normalfont\AA}$ 			\\
effective spin per unit cell 									& $S$ 							& $20$										& 															\\
\phantom{a} for $T\lesssim 50\,\textrm{K}$ 		&										&													&																\\
exchange stiffness constant 									& $J_s$ 						& $8.458\times 10^{-40}$ 	& $\textrm{J}\,\textrm{m}^2$ 		\\
$g$-factor																		& $g$								& $2$											&																\\
mass per unit cell 														& $m$ 							& $9.800\times 10^{-24}$ 	& $\textrm{kg}$ 								\\
longitudinal sound velocity 									& $c_\parallel$ 		& $7209$ 									& $\textrm{m}\,\textrm{s}^{-1}$ \\
transverse sound velocity 										& $c_\bot$ 					& $3843$ 									& $\textrm{m}\,\textrm{s}^{-1}$ \\
longitudinal critical field										& $B_{c,\parallel}$	&	$9.21$									&	$\textrm{T}$									\\
transverse critical field											& $B_{c,\bot}$			&	$2.62$									&	$\textrm{T}$									\\
magnetic Gr\"{u}neisen												& $\Gamma_m$				& $-3.2$									&																\\	
\phantom{a} parameter													&										&													&																\\
diagonal magnetoelastic 											& $B_\parallel$ 		& $6.597\times 10^{-22}$ 	& $\textrm{J}$ 									\\
\phantom{a} constant													&										&													&																\\
off-diagonal magnetoelastic 									& $B_\bot$ 					& $1.319\times 10^{-21}$ 	& $\textrm{J}$ 									\\
\phantom{a} constant													&										&													&																\\
exchange magnetoelastic 											& $A_\parallel$ 		& $-8.120\times 10^{-38}$ & $\textrm{J}\,\textrm{m}^2$ 		\\
\phantom{a} constant													&										&													&																\\
\hline\hline
\end{tabular}
\label{tab:YIG_values} 
\end{table}
We discuss three scenarios: (i) heating, (ii) parametric pumping by
microwaves, and (iii) optical spin injection. First, we consider the scenario
in which energy injected into the lattice, e.g., by a femtosecond laser pulse
at an optical phonon resonance, relaxes very quickly to a distribution of the
form (\ref{eq:N_ansatz}), which subsequently releases energy to the magnetic
system. In this case there is no angular momentum transfer from the
environment and $\delta j_{0}=0$. Figure \ref{figHeating} shows the calculated
dynamics when the magnetic order is perturbed by a sudden increase of the
phonon temperature $\delta T_{\parallel}(0)=\delta T_{\bot}(0)$.

\begin{figure*}
\includegraphics[width=.82\textwidth]{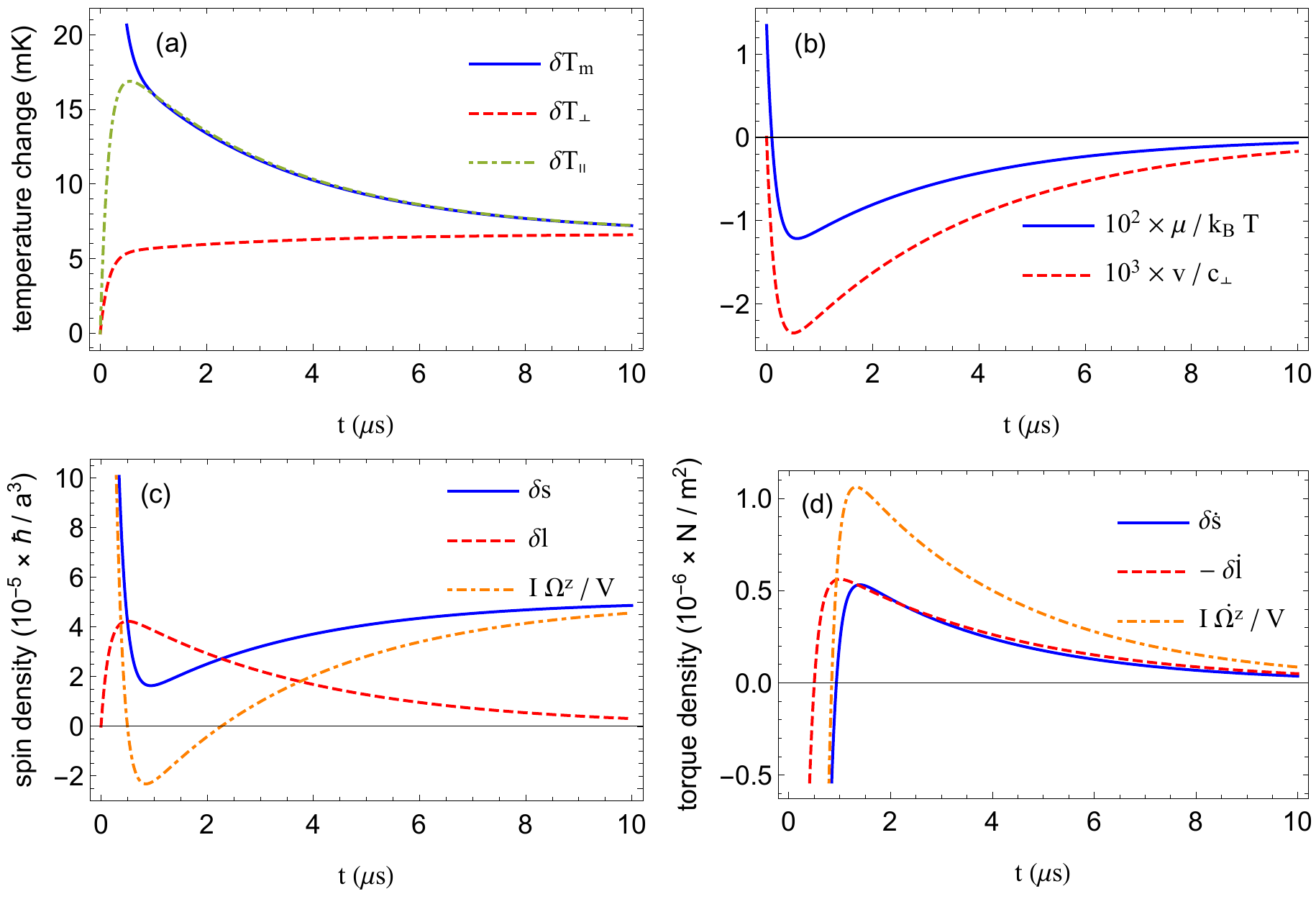}
\caption{\label{figPumping} Nonequilibrium dynamics of 
(a) magnon and phonon temperatures,
(b) magnon chemical potential and phonon drift velocity,
(c) spin densities, and 
(d) torque densities 
of magnons, phonons, and rigid rotation,
with initial conditions
$\delta T_m(t=0)=1\,\textrm{K}$, $\mu(t=0)=0.1\times \hbar \gamma B$,
and $\delta T_\bot(t=0)=\delta T_\parallel(t=0)=v(t=0)=0$,
corresponding to magnon pumping, e.g., by applying a parallel parametric pumping field.
Temperature and external magnetic field are $T=10\,\textrm{K}$ and $B=1\,\textrm{T}$.}
\end{figure*}

\begin{figure*}
\includegraphics[width=.82\textwidth]{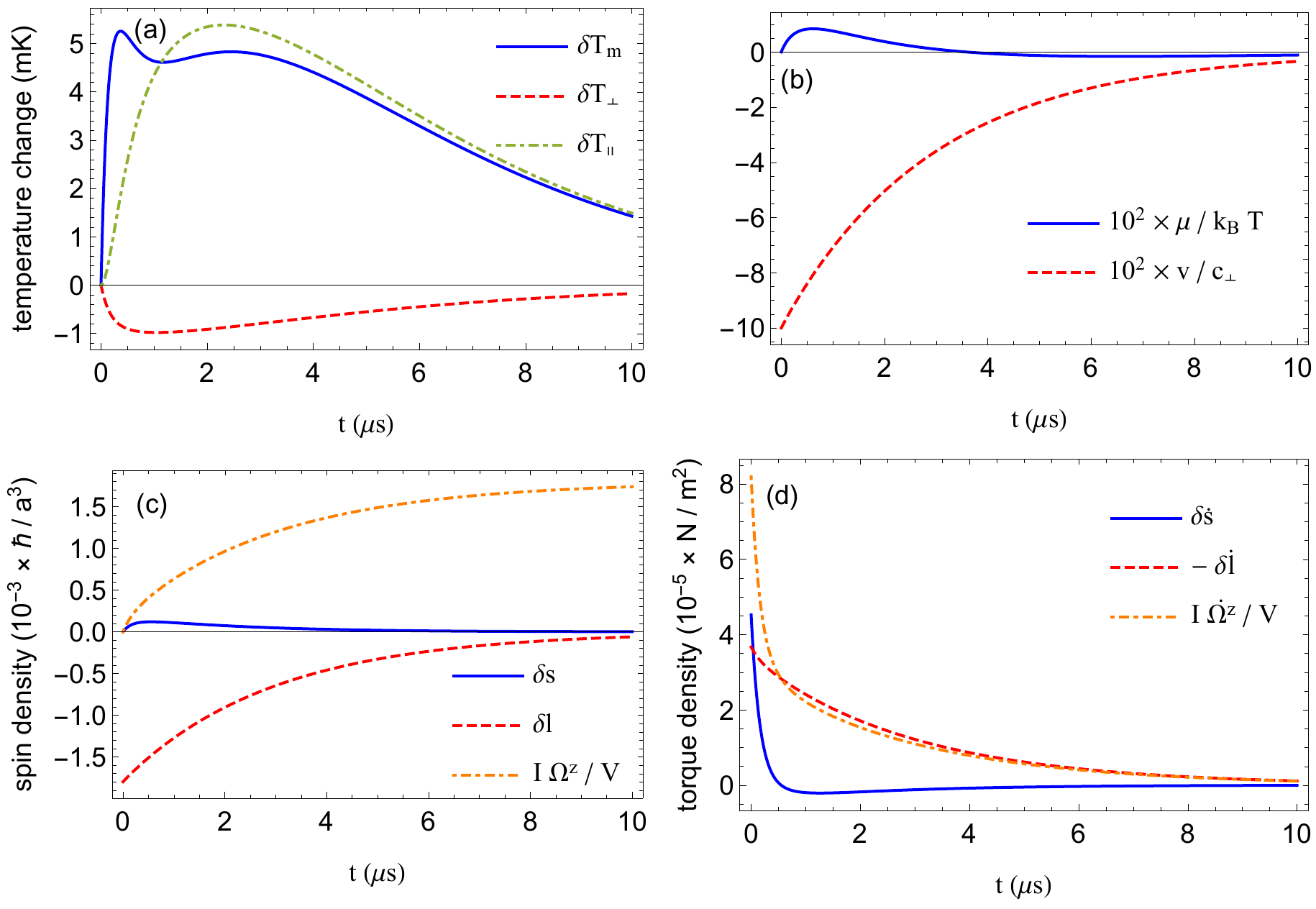}
\caption{\label{figPhonon} Nonequilibrium dynamics of 
(a) magnon and phonon temperatures,
(b) magnon chemical potential and phonon drift velocity,
(c) spin densities, and 
(d) torque densities 
of magnons, phonons, and rigid rotation,
with initial conditions
$v(t=0)=-0.1\times c_\bot$,
and $\delta T_\bot(t=0)=\delta T_\parallel(t=0)=\delta T_m(t=0)=\mu(t=0)=0$.
corresponding to a finite phonon spin.
Temperature and external magnetic field are $T=10\,\textrm{K}$ and $B=1\,\textrm{T}$.}
\end{figure*}

Parallel microwave pumping is the nonlinear process in which a
microwave magnetic field parallel to the magnetization parametrically excites
the Kittel mode above a certain threshold intensity. In contrast to the
(linear) ferromagnetic resonance, the linearly polarized radiation does not
inject angular momentum into the magnet, so also in this case $\delta
j_{0}=0.$ The angular momentum needed to excite the magnetization is therefore
provided only by the lattice. We assume that the pumped magnons thermalize
quickly to a distribution with increased temperature and finite magnon
chemical potential, while the lattice is initially in equilibrium and plot the
results in Fig. \ref{figPumping}.
The third scenario addresses the direct injection of angular
momentum into the phonons. This can be achieved by exposing the magnet to
circularly polarized light that couples only to phonons with a certain spin
polarization, or by phonon spin pumping from a thin film of another magnet
attached to the system \cite{Streib2018}. Since the phonon spin is supplied by
the external environment in this case, we have $\delta j_{0}=\delta l(t=0)$,
while the magnons are initially in equilibrium. The response to such an
external torque is plotted in Fig. \ref{figPhonon}.

The cases (i) and (ii) share many features. Figures \ref{figHeating}(a) and
\ref{figPumping}(a) show that the energy relaxes in two stages: First, the
longitudinal-phonon and magnon temperatures converge, after which they both
equilibrate with the transverse-phonon temperature. The longitudinal phonons
and the magnons equilibrate faster than the transverse phonons, because the
specific heat of the former is an order of magnitude smaller than that of the
latter. In parallel, the magnetoelastic coupling builds up transient,
counterpropagating currents of the two circular phonon modes, i.e., a phononic
spin, on a time scale similar to the magnon chemical potential or spin
accumulation, see Figs.~\ref{figHeating}(b) and \ref{figPumping}(b). The
phonon spin density generated by phonon heating is typically an order of
magnitude smaller than the magnon spin. The induced rigid rotation in
Fig.~\ref{figHeating}(c) is therefore mainly a magnonic effect. However, when
the system is excited by pumping the magnons, the phonon spin transiently
dominates the magnon contribution, see Fig.~\ref{figPumping}(c). The angular
velocity $\Omega^{z}$ temporarily changes sign, i.e., the body rotates in the 
opposite direction, seemingly breaking the angular conservation law. After the
magnon-dominated first microsecond, the torques exerted by both phonons and
magnons in Figs.~\ref{figHeating}(d) and \ref{figPumping}(d) are very similar.

Figure \ref{figPhonon} sketches the even more dramatic effect when the injected
phonons initially carry a spin without excess energy, which means that the
system at large times must relax to the initial temperature $T.$ However, spin
may be transferred from the phonons to the magnons, which heats the magnons
and endows them with a finite chemical potential, see Figs.~\ref{figPhonon}%
(a)-\ref{figPhonon}(c), which is only possible by transient cooling of the transverse
phonons. Actually only a small fraction of the spin is transferred from the
phonons to the magnons in Fig.~\ref{figPhonon}(c): The loss of phonon spin is
accommodated by the rigid rotation of the entire magnet. The overall torque in
Fig.~\ref{figPhonon}(d) is dominated by the phonons at almost all times.

\section{Discussion and Conclusions}
\label{sec:conclusions}

We present a microscopic theory of spin-lattice interactions and angular
momentum conservation in magnetic insulators. After separating the mechanical
degrees of freedom into rigid-body and internal vibrations, we find that also
phonons carry internal angular momentum. We derive equations of motion for the
spin, rigid-body, and phonon spin operators that govern the Einstein-de Haas
and Barnett effects, and show that the torque generated by spin-lattice
interactions drives both the rigid-body rotation and the phonon spin. In the
long-wavelength limit, we recover the phenomenological theory of magnetoelasticity.

We apply the formalism to a linear response analysis of a levitated magnet
that is large enough that surface effects can be disregarded, but small enough
that rotations are observable. In contrast to the magnon chemical potential or
accumulation, the phonon chemical potential does not couple to the total
rotation. It is rather an internal phonon current that governs the phonon
contribution to the Einstein-de Haas effect. Depending on the driving
protocol, the transient Einstein-de Haas dynamics can involve a change in the
sense of rotation. When the system is not levitated but fixed, e.g., on a
substrate, the torques exerted by the magnon and phonon spins on the sample
are in principle measurable \cite{Wallis2006,Zolfagharkhani2008,Harii2019}.
Brillouin light scattering \cite{Holanda2018} can resolve the phonon spin; our
prediction of a momentum imbalance between the two circularly polarized phonon
modes should therefore also be experimentally accessible.

Several assumptions and approximations imply that the present results are
valid for a limited temperature and size of the system. The adoption of
the magnetoelastic limit implies that temperatures should not exceed the
frequencies for which a continuum mechanics and magnetism holds, roughly
$T<100\,$K. The decoupling of internal phonon modes from the total rotation
introduces errors that we estimate to vanish when the number of spins is much
larger than unity which is not very restrictive. More drastic is the
assumption that the phonon relaxation length should be much smaller than the
systems size, in order to allow the flow of transient phonon spin currents.
This is a material specific and temperature dependent parameter that is not
well known. When the phonon relaxation length is much larger than the particle
size, a phonon spin does not build up, strongly suppressing the phonon contribution to the Einstein-de
Haas effect. For materials with extremely low acoustic attenuation such as
YIG, the phonon propagation length at GHz frequencies can be centimeters.
Thermal phonons at not too low temperatures are more strongly scattered, which
leads us to believe that YIG spheres that can be fabricated for diameters
$\gtrsim0.5$ mm are suitable model systems to test our predictions. The size
estimates for other materials can be substantially smaller, however. For
particles larger than the phonon relaxation length, the ratio between
predicted torques and total volume is predicted to be constant as long as the
excitation is more or less homogeneous.     

Our treatment of angular momentum transfer in spin-lattice interactions should
be useful in the study of a variety of problems. Of particular interest would
be the application to the magnetic nanosystems like cantilevers
\cite{Wallis2006,Zolfagharkhani2008,Harii2019} and nanoparticles in polymer
cavities \cite{Tejada2010} or levitated in traps
\cite{Rusconi2016,Prat-Camps2017,Rusconi2017,Ballestero2019}. It could also be
extended to study the role of the phonon spin in transport phenomena like the
spin Seebeck effect \cite{Uchida2010}. Moreover, the microscopic spin-lattice
Hamiltonian that we proposed could be used to extend computations of
magnon-phonon interactions \cite{Streib2019} into the high-temperature regime
where magnetoelastic theory is no longer valid, and to determine material
constants from \textit{ab initio} computations. Another extension of the
formalism would allow addressing finite-size corrections, quantum effects, and
time scales at which rigid-body and internal phonon dynamics cannot be separated.

\begin{acknowledgments}
This work is supported by the European Research Council via Consolidator Grant No.~725509 SPINBEYOND
and JSPS KAKENHI Grant No.~19H006450. 
R.D. is a member of the D-ITP consortium, a program of the Netherlands Organisation for Scientific Research (NWO) that is funded by the Dutch Ministry of Education, Culture and Science (OCW).
This work is also part of the normal research program of NWO.
This research was supported in part by the National Science
Foundation under Grant No.~NSF PHY-1748958.
\end{acknowledgments}

\appendix

\section{Total Angular Momentum Conservation}
\label{app:J}

In terms of the three Euler angles $\phi$, $\theta$, and $\chi$ as defined in
Eq.~(\ref{eq:r_expansion}) the body-fixed total angular momentum operator
reads (see, e.g., Refs.~\cite{Louck1976,Littlejohn1997} for details)
\begin{equation} \label{eq:J_Euler}
\bm{J} = \frac{\hbar}{i}
\left( \begin{matrix}
\partial_\chi \\
\frac{\sin\chi}{\cos\theta} \partial_\phi + \cos\chi \partial_\theta + \tan\theta \sin\chi \partial_\chi \\
\frac{\cos\chi}{\cos\theta} \partial_\phi - \sin\chi \partial_\theta + \tan\theta \cos\chi \partial_\chi 
\end{matrix} \right) .
\end{equation}
In the laboratory frame
\begin{align} \label{eq:J_Euler_lab}
\bm{J}_{\textrm{lab}} 
&= {\cal R}(\phi,\theta,\chi)\bm{J} 
\\
&= \frac{\hbar}{i}
\left( \begin{matrix}
\cos\phi\tan\theta \partial_\phi - \sin\phi \partial_\theta + \frac{\cos\phi}{\cos\theta} \partial_\chi \\
\sin\phi\tan\theta \partial_\phi + \cos\phi \partial_\theta + \frac{\sin\phi}{\cos\theta} \partial_\chi \\
\partial_\phi
\end{matrix} \right) .
\end{align}
These operators obey the commutation relations
\begin{align}
\left[ J^{x}, J^{y} \right] =& - i\hbar J^{z} , \\
\left[ J^x_{\textrm{lab}}, J^y_{\textrm{lab}} \right] =& i\hbar J^z_{\textrm{lab}} ,
\end{align}
including their cyclic permutations. 
Also, $\bm{J}^{2}=\bm{J}_{\mathrm{lab}}^{2}$, and
\begin{align}
\left[ \bm{J}^2_{\textrm{lab}} , \bm{J} \right] =& 0 , \\
\left[ J^z_{\textrm{lab}} , \bm{J} \right] =& 0 ,
\end{align}
whereas
\begin{align}
\left[ \bm{J}_{\textrm{lab}} , U^\dagger H_S U \right] =
%-{\cal R}^T(\phi,\theta,\chi)\bm{h} \cdot\sum_{i=1}^N \bm{S}_i \right] =
&
i\hbar \sum_{i=1}^N \left[ {\cal R}(\phi,\theta,\chi) \bm{S}_i \right] \times \gamma\bm{B} 
\nonumber\\
&
+ i\hbar {\cal R}(\phi,\theta,\chi) \bm{{\cal T}}_{\rm ext}.
\end{align}
Hence, in the absence of torques by an external magnetic field $\bm{B}$ or
external mechanical forces, the absolute value $\bm{J}_{\mathrm{lab}}^{2}$ and
the $z$ component $J_{\mathrm{lab}}^{z}$ of the total angular momentum are conserved.

\section{Phonon Commutation Relations}

\label{sec:Commutation}

The commutator of the phonon displacement and position operators introduced in
Eq.~(\ref{eq:def_u_pi}) is given by
\begin{align}
\left[  u_{i}^{\alpha},\pi_{j}^{\beta}\right]  = 
&  i\hbar\sqrt{\frac{m_{j}}{m_{i}}}\sum_{n=1}^{3N-6}f_{n}^{\alpha}(\bm{R}_{i})f_{m}^{\beta}(\bm{R}_{j})
\\
= 
&  i\hbar\biggl[  \delta_{ij}\delta^{\alpha\beta}  
\nonumber\\
& -\sqrt{\frac{m_{j}}{m_{i}}}\sum_{n\in\mathrm{zero\ modes}}
f_{n}^{\alpha}(\bm{R}_{i})f_{m}^{\beta}(\bm{R}_{j})\biggr] ,
\label{eq:commutator_explicit}
\end{align}
where we used the completeness relation $\sum_{n=1}^{3N}f_{n}^{\alpha
}(\bm{R}_{i})f_{m}^{\beta}(\bm{R}_{j})=\delta_{ij}\delta^{\alpha\beta}$ for
the combined phonon and rigid-body zero modes.

Explicit expressions for the $6$ zero mode eigenfunctions of center-of-mass
translation and rigid rotation must obey the translational and rotational
invariance. The transformation $\bm{R}_{i}\rightarrow\bm{R}_{i}+\bm{a}$, where
$\bm{a}$ is a constant vector, leaves the potential $V(\{\bm{r}_{i}\})$
invariant. Comparing to the definition (\ref{eq:r_expansion}) of the phonon
eigenmode expansion in the body-fixed frame, we find the $3$ (normalized) zero
mode eigenfunctions of center-of-mass translation
\begin{equation}
\bm{f}_{CM,\mu}(\bm{R}_{i})=\sqrt{\frac{m_{i}}{M}}\bm{\hat{e}}_{\mu},
\end{equation}
where the $\bm{\hat{e}}_{\mu}$ with $\mu=1,2,3$ are an arbitrary set of
orthonormal basis vectors. Similarly, we obtain the $3$ zero modes of rigid
rotation by considering infinitesimal rotations $\bm{R}_{i}\rightarrow
\bm{R}_{i}+\bm{\phi}\times\bm{R}_{i}$, with $\left\vert \bm{\phi}\right\vert
\ll1$, yielding
\begin{equation}
\bm{f}_{R,\mu}(\bm{R}_{i})=\sqrt{\frac{m_{i}}{I_{\mu}}}\bm{\hat{n}}_{\mu
}\times\bm{R}_{i}.
\end{equation}
Here the $\bm{\hat{n}}_{\mu}$ with $\mu=1,2,3$ are the principal axes of the
system in the body-fixed frame, and $I_{\mu}$ the corresponding principal
moments of inertia. The commutator (\ref{eq:commutator_explicit}) becomes
\begin{align}
\left[  u_{i}^{\alpha},\pi_{j}^{\beta}\right]  = &  i\hbar\Biggl[  
\delta_{ij}\delta^{\alpha\beta}-\frac{m_{j}}{M}\delta^{\alpha\beta}
\nonumber\\
&  
-\sum_{\mu=1}^{3}\frac{m_{j}\left(  \bm{\hat{n}}_{\mu}
\times\bm{R}_{i}\right)  ^{\alpha}\left(  \bm{\hat{n}}_{\mu}\times
\bm{R}_{j}\right)  ^{\beta}}{I_{\mu}}\Biggr]  .
\end{align}

We can estimate the order of magnitude of the corrections to the commutation
relations by introducing a unit cell around each particle with volume $\Delta
V=V/N$ and mass $m_{i}\approx\rho\Delta V$, where $\rho=M/V$ is the mass
density. The maximum distance $\bm{R}_{i}$ of a particle from the origin is of
the order $V^{1/3}$, so the moment of inertia $I_{\mu}\sim\rho V^{5/3}$.
Hence
\begin{subequations}
\begin{align}
\frac{m_{j}}{M} &  \sim\frac{\rho\Delta V}{\rho V}=\frac{1}{N},\\
\frac{m_{j}\left(  \bm{\hat{n}}_{\mu}\times\bm{R}_{i}\right)  ^{\alpha}\left(
\bm{\hat{n}}_{\mu}\times\bm{R}_{j}\right)  ^{\beta}}{I_{\mu}} &  \sim
\frac{\rho\Delta VV^{2/3}}{\rho V^{5/3}}=\frac{1}{N}.
\end{align}
\end{subequations}
Therefore, the noncanonical corrections to the commutator
(\ref{eq:commutator_explicit}) scale with the inverse of the number of
particles $N$ in the system; hence the phonon operators (\ref{eq:def_u_pi})
and (\ref{eq:L}) in the body-fixed frame can be treated as canonical whenever
$N\gg1$, so one has to worry about corrections only for small molecules.

\section{Linear Response Relaxation Rates}
\label{sec:Rates}

The linear response relaxation rates
\begin{equation}
\Gamma_{\alpha\beta}=\Gamma_{\alpha\beta}^{(1)}+\Gamma_{\alpha\beta}^{(2)},
\end{equation}
where $\Gamma_{\alpha\beta}^{(1)}$ and $\Gamma_{\alpha\beta}^{(2)}$ are due to
one-magnon one-phonon and two-magnon one-phonon processes respectively, follow
from inserting the ans\"{a}tze (\ref{eq:n_ansatz}) and (\ref{eq:N_ansatz}) for
the magnon and phonon distribution functions into the kinetic equations
(\ref{eq:ndot}) and (\ref{eq:Ndot}). Explicitly,
\begin{widetext}
\begin{subequations} \label{eq:Gamma1}
\begin{align}
\Gamma_{\bot}^{(1)} =
& 
\frac{\pi \hbar}{m k_B^2 T^2 V} \sum_{\bm{k}} \omega_{\bm{k}\lambda} \delta_{\lambda,\bot}
| \bm{\hat{e}}_{\bm{k}\lambda}^* \cdot \bm{\Gamma}_{\bm{k}} |^2 \delta( \epsilon_{\bm{k}} - \omega_{\bm{k}\lambda} )
\left[ 1 + f_B\left( \frac{\hbar \epsilon_{\bm{k}} }{ k_B T } \right) \right] 
f_B\left( \frac{\hbar \epsilon_{\bm{k}} }{ k_B T } \right) ,
\\[.25cm]
\Gamma_{\parallel}^{(1)} =
& 
\frac{\pi \hbar}{m k_B^2 T^2 V} \sum_{\bm{k}} \omega_{\bm{k}\lambda} \delta_{\lambda,\parallel}
| \bm{\hat{e}}_{\bm{k}\lambda}^* \cdot \bm{\Gamma}_{\bm{k}} |^2 \delta( \epsilon_{\bm{k}} - \omega_{\bm{k}\lambda} )
\left[ 1 + f_B\left( \frac{\hbar \epsilon_{\bm{k}} }{ k_B T } \right) \right] 
f_B\left( \frac{\hbar \epsilon_{\bm{k}} }{ k_B T } \right) ,
\\[.25cm]
\Gamma_{v}^{(1)} =
& \frac{\pi \hbar c_\bot^2}{m k_B^2 T^2 V} \sum_{\bm{k}\lambda} \delta_{\lambda,\bot} \left( k^z \right)^2 
\frac{ | \bm{\hat{e}}_{\bm{k}\lambda}^* \cdot \bm{\Gamma}_{\bm{k}} |^2 }{ \omega_{\bm{k}\lambda} }
\delta( \epsilon_{\bm{k}} - \omega_{\bm{k}\lambda} )
\left[ 1 + f_B\left( \frac{\hbar \epsilon_{\bm{k}} }{ k_B T } \right) \right] 
f_B\left( \frac{\hbar \epsilon_{\bm{k}} }{ k_B T } \right) ,
\\[.25cm]
\Gamma_{\bot v}^{(1)} =
& \frac{\pi \hbar c_\bot}{m k_B^2 T^2 V} \sum_{\bm{k}\lambda} \delta_{\lambda,\bot} \lambda k^z
| \bm{\hat{e}}_{\bm{k}\lambda}^* \cdot \bm{\Gamma}_{\bm{k}} |^2 \delta( \epsilon_{\bm{k}} - \omega_{\bm{k}\lambda} )
\left[ 1 + f_B\left( \frac{\hbar \epsilon_{\bm{k}} }{ k_B T } \right) \right] 
f_B\left( \frac{\hbar \epsilon_{\bm{k}} }{ k_B T } \right) ,
\\[.25cm]
\Gamma_{\bot\mu}^{(1)} =
& - \frac{\pi}{m k_B T V} \sum_{\bm{k}\lambda} \delta_{\lambda,\bot} 
| \bm{\hat{e}}_{\bm{k}\lambda}^* \cdot \bm{\Gamma}_{\bm{k}} |^2 \delta( \epsilon_{\bm{k}} - \omega_{\bm{k}\lambda} )
\left[ 1 + f_B\left( \frac{\hbar \epsilon_{\bm{k}} }{ k_B T } \right) \right] 
f_B\left( \frac{\hbar \epsilon_{\bm{k}} }{ k_B T } \right) ,
\\[.25cm]
\Gamma_{\parallel\mu}^{(1)} =
& - \frac{\pi}{m k_B T V} \sum_{\bm{k}\lambda} \delta_{\lambda,\parallel} 
| \bm{\hat{e}}_{\bm{k}\lambda}^* \cdot \bm{\Gamma}_{\bm{k}} |^2 \delta( \epsilon_{\bm{k}} - \omega_{\bm{k}\lambda} )
\left[ 1 + f_B\left( \frac{\hbar \epsilon_{\bm{k}} }{ k_B T } \right) \right] 
f_B\left( \frac{\hbar \epsilon_{\bm{k}} }{ k_B T } \right) ,
\\[.25cm]
\Gamma_{v \mu}^{(1)} =
& - \frac{\pi c_\bot}{m k_B T V} \sum_{\bm{k}\lambda} \delta_{\lambda,\bot} \lambda k^z 
\frac{ | \bm{\hat{e}}_{\bm{k}\lambda}^* \cdot \bm{\Gamma}_{\bm{k}} |^2 }{ \omega_{\bm{k}\lambda} }
\delta( \epsilon_{\bm{k}} - \omega_{\bm{k}\lambda} )
\left[ 1 + f_B\left( \frac{\hbar \epsilon_{\bm{k}} }{ k_B T } \right) \right] 
f_B\left( \frac{\hbar \epsilon_{\bm{k}} }{ k_B T } \right) ,
\\[.25cm]
\Gamma_{\mu}^{(1)} =
& \frac{\pi}{\hbar m V} \sum_{\bm{k}\lambda} 
\frac{ | \bm{\hat{e}}_{\bm{k}\lambda}^* \cdot \bm{\Gamma}_{\bm{k}} |^2 }{ \omega_{\bm{k}\lambda} }
\delta( \epsilon_{\bm{k}} - \omega_{\bm{k}\lambda} )
\left[ 1 + f_B\left( \frac{\hbar \epsilon_{\bm{k}} }{ k_B T } \right) \right] 
f_B\left( \frac{\hbar \epsilon_{\bm{k}} }{ k_B T } \right) ,
\end{align}
\end{subequations}
and
\begin{subequations} \label{eq:Gamma2}
\begin{align}
\Gamma_{\bot}^{(2)} =
&
\frac{ \pi \hbar a^3 }{ m k_B^2 T^2 V^2 } \sum_{\bm{k}\bm{k}'\bm{q}\lambda}
\delta_{\lambda,\bot} \omega_{\bm{q}\lambda} \Biggl\{
\delta_{\bm{k}-\bm{k}',\bm{q}}
| \bm{\hat{e}}_{\bm{q}\lambda} \cdot \bm{U}_{\bm{k},\bm{k}'} |^2
\delta( \epsilon_{\bm{k}} - \epsilon_{\bm{k}'} - \omega_{\bm{q}\lambda} )
\left[ 1 + f_B\left( \frac{\hbar \epsilon_{\bm{k}} }{ k_B T } \right) \right] 
f_B\left( \frac{\hbar \epsilon_{\bm{k}'} }{ k_B T } \right)
f_B\left( \frac{\hbar \omega_{\bm{q}\lambda} }{ k_B T } \right)
\nonumber\\
&
+
\delta_{\bm{k}+\bm{k}',\bm{q}}
\frac{ | \bm{\hat{e}}_{\bm{q}\lambda}^* \cdot \bm{V}_{\bm{k},\bm{k}'} |^2 }{ 2 }
\delta( \epsilon_{\bm{k}} + \epsilon_{\bm{k}'} - \omega_{\bm{q}\lambda} )
\left[ 1 + f_B\left( \frac{\hbar \epsilon_{\bm{k}} }{ k_B T } \right) \right] 
\left[ 1 + f_B\left( \frac{\hbar \epsilon_{\bm{k}'} }{ k_B T } \right) \right]
f_B\left( \frac{\hbar \omega_{\bm{q}\lambda} }{ k_B T } \right)
\Biggr\}, 
\\[.25cm]
\Gamma_{\parallel}^{(2)} =
&
\frac{ \pi \hbar a^3 }{ m k_B^2 T^2 V^2 } \sum_{\bm{k}\bm{k}'\bm{q}\lambda}
\delta_{\lambda,\parallel} \omega_{\bm{q}\lambda} \Biggl\{
\delta_{\bm{k}-\bm{k}',\bm{q}}
| \bm{\hat{e}}_{\bm{q}\lambda} \cdot \bm{U}_{\bm{k},\bm{k}'} |^2
\delta( \epsilon_{\bm{k}} - \epsilon_{\bm{k}'} - \omega_{\bm{q}\lambda} )
\left[ 1 + f_B\left( \frac{\hbar \epsilon_{\bm{k}} }{ k_B T } \right) \right] 
f_B\left( \frac{\hbar \epsilon_{\bm{k}'} }{ k_B T } \right)
f_B\left( \frac{\hbar \omega_{\bm{q}\lambda} }{ k_B T } \right)
\nonumber\\
&
+
\delta_{\bm{k}+\bm{k}',\bm{q}}
\frac{ | \bm{\hat{e}}_{\bm{q}\lambda}^* \cdot \bm{V}_{\bm{k},\bm{k}'} |^2 }{ 2 }
\delta( \epsilon_{\bm{k}} + \epsilon_{\bm{k}'} - \omega_{\bm{q}\lambda} )
\left[ 1 + f_B\left( \frac{\hbar \epsilon_{\bm{k}} }{ k_B T } \right) \right] 
\left[ 1 + f_B\left( \frac{\hbar \epsilon_{\bm{k}'} }{ k_B T } \right) \right]
f_B\left( \frac{\hbar \omega_{\bm{q}\lambda} }{ k_B T } \right)
\Biggr\}, 
\\[.25cm]
\Gamma_{v}^{(2)} =
&
\frac{ \pi \hbar a^3 c_\bot^2 }{ m k_B^2 T^2 V^2 } 
\sum_{\bm{k}\bm{k}'\bm{q}\lambda} \delta_{\lambda,\bot} \left( q^z \right)^2
\Biggl\{
 \delta_{\bm{k}-\bm{k}',\bm{q}}
\frac{ | \bm{\hat{e}}_{\bm{q}\lambda} \cdot \bm{U}_{\bm{k},\bm{k}'} |^2 }{ \omega_{\bm{q}\lambda} }
\delta( \epsilon_{\bm{k}} - \epsilon_{\bm{k}'} - \omega_{\bm{q}\lambda} )
\left[ 1 + f_B\left( \frac{\hbar \epsilon_{\bm{k}} }{ k_B T } \right) \right] 
f_B\left( \frac{\hbar \epsilon_{\bm{k}'} }{ k_B T } \right)
f_B\left( \frac{\hbar \omega_{\bm{q}\lambda} }{ k_B T } \right)
\nonumber\\
&
+
\delta_{\bm{k}+\bm{k}',\bm{q}}
\frac{ | \bm{\hat{e}}_{\bm{q}\lambda}^* \cdot \bm{V}_{\bm{k},\bm{k}'} |^2 }{ 2 \omega_{\bm{q}\lambda} }
\delta( \epsilon_{\bm{k}} + \epsilon_{\bm{k}'} - \omega_{\bm{q}\lambda} )
\left[ 1 + f_B\left( \frac{\hbar \epsilon_{\bm{k}} }{ k_B T } \right) \right] 
\left[ 1 + f_B\left( \frac{\hbar \epsilon_{\bm{k}'} }{ k_B T } \right) \right]
f_B\left( \frac{\hbar \omega_{\bm{q}\lambda} }{ k_B T } \right)
\Biggr\}, 
\\[.25cm]
\Gamma_{\bot v}^{(2)} =
&
\frac{ \pi \hbar a^3 c_\bot }{ m k_B^2 T^2 V^2 } 
\sum_{\bm{k}\bm{k}'\bm{q}\lambda} \delta_{\lambda,\bot} \lambda q^z 
\Biggl\{
\delta_{\bm{k}-\bm{k}',\bm{q}} 
| \bm{\hat{e}}_{\bm{q}\lambda} \cdot \bm{U}_{\bm{k},\bm{k}'} |^2
\delta( \epsilon_{\bm{k}} - \epsilon_{\bm{k}'} - \omega_{\bm{q}\lambda} )
\left[ 1 + f_B\left( \frac{\hbar \epsilon_{\bm{k}} }{ k_B T } \right) \right] 
f_B\left( \frac{\hbar \epsilon_{\bm{k}'} }{ k_B T } \right)
f_B\left( \frac{\hbar \omega_{\bm{q}\lambda} }{ k_B T } \right)
\nonumber\\
&
+
\delta_{\bm{k}+\bm{k}',\bm{q}}
\frac{ | \bm{\hat{e}}_{\bm{q}\lambda}^* \cdot \bm{V}_{\bm{k},\bm{k}'} |^2 }{ 2 }
\delta( \epsilon_{\bm{k}} + \epsilon_{\bm{k}'} - \omega_{\bm{q}\lambda} )
\left[ 1 + f_B\left( \frac{\hbar \epsilon_{\bm{k}} }{ k_B T } \right) \right] 
\left[ 1 + f_B\left( \frac{\hbar \epsilon_{\bm{k}'} }{ k_B T } \right) \right]
f_B\left( \frac{\hbar \omega_{\bm{q}\lambda} }{ k_B T } \right)
\Biggr\}, 
\\[.25cm]
\Gamma_{\bot\mu}^{(2)} =
&
- \frac{ \pi a^3 }{ m k_B T V^2 } \sum_{\bm{k}\bm{k}'\bm{q}\lambda} \delta_{\lambda,\bot} \delta_{\bm{k}+\bm{k}',\bm{q}}
| \bm{\hat{e}}_{\bm{q}\lambda}^* \cdot \bm{V}_{\bm{k},\bm{k}'} |^2 
\delta( \epsilon_{\bm{k}} + \epsilon_{\bm{k}'} - \omega_{\bm{q}\lambda} )
\left[ 1 + f_B\left( \frac{\hbar \epsilon_{\bm{k}} }{ k_B T } \right) \right] 
\left[ 1 + f_B\left( \frac{\hbar \epsilon_{\bm{k}'} }{ k_B T } \right) \right]
f_B\left( \frac{\hbar \omega_{\bm{q}\lambda} }{ k_B T } \right), 
\\[.25cm]
\Gamma_{\parallel\mu}^{(2)} =
&
- \frac{ \pi a^3 }{ m k_B T V^2 } \sum_{\bm{k}\bm{k}'\bm{q}\lambda} \delta_{\lambda,\parallel} \delta_{\bm{k}+\bm{k}',\bm{q}}
| \bm{\hat{e}}_{\bm{q}\lambda}^* \cdot \bm{V}_{\bm{k},\bm{k}'} |^2 
\delta( \epsilon_{\bm{k}} + \epsilon_{\bm{k}'} - \omega_{\bm{q}\lambda} )
\left[ 1 + f_B\left( \frac{\hbar \epsilon_{\bm{k}} }{ k_B T } \right) \right] 
\left[ 1 + f_B\left( \frac{\hbar \epsilon_{\bm{k}'} }{ k_B T } \right) \right]
f_B\left( \frac{\hbar \omega_{\bm{q}\lambda} }{ k_B T } \right), 
\\[.25cm]
\Gamma_{v \mu}^{(2)} =
&
- \frac{ \pi a^3 c_\bot }{ m k_B T V^2 }
\sum_{\bm{k}\bm{k}'\bm{q}\lambda} \delta_{\lambda,\bot} \delta_{\bm{k}+\bm{k}',\bm{q}}
\lambda q^z \frac{ | \bm{\hat{e}}_{\bm{q}\lambda}^* \cdot \bm{V}_{\bm{k},\bm{k}'} |^2 }{ \omega_{\bm{q}\lambda} }
\delta( \epsilon_{\bm{k}} + \epsilon_{\bm{k}'} - \omega_{\bm{q}\lambda} )
\nonumber\\
&
\phantom{ \frac{ \pi \hbar a^3 }{ 2 m k_B T^2 V } \sum_{\bm{k}\bm{k}'\bm{q}\lambda} } \times
\left[ 1 + f_B\left( \frac{\hbar \epsilon_{\bm{k}} }{ k_B T } \right) \right] 
\left[ 1 + f_B\left( \frac{\hbar \epsilon_{\bm{k}'} }{ k_B T } \right) \right]
f_B\left( \frac{\hbar \omega_{\bm{q}\lambda} }{ k_B T } \right), 
\\[.25cm]
\Gamma_{\mu}^{(2)} =
&
\frac{ 2 \pi a^3 }{ \hbar m V^2 }
\sum_{\bm{k}\bm{k}'\bm{q}\lambda} \delta_{\bm{k}+\bm{k}',\bm{q}}
\frac{ | \bm{\hat{e}}_{\bm{q}\lambda}^* \cdot \bm{V}_{\bm{k},\bm{k}'} |^2 }{ \omega_{\bm{q}\lambda} }
\delta( \epsilon_{\bm{k}} + \epsilon_{\bm{k}'} - \omega_{\bm{q}\lambda} )
\left[ 1 + f_B\left( \frac{\hbar \epsilon_{\bm{k}} }{ k_B T } \right) \right] 
\left[ 1 + f_B\left( \frac{\hbar \epsilon_{\bm{k}'} }{ k_B T } \right) \right]
f_B\left( \frac{\hbar \omega_{\bm{q}\lambda} }{ k_B T } \right), 
\end{align}
\end{subequations}
\end{widetext}
where 
$\delta_{\lambda,\bot}=\delta_{\lambda,+}+\delta_{\lambda,-}$.

\end{document}